\documentclass[useAMS,usenatbib]{mnras}
\usepackage{amsmath}
\usepackage{amssymb}
\usepackage{graphicx}
\usepackage{verbatim}
\pagerange{\pageref{firstpage}--\pageref{lastpage}}
\pubyear{2020}
\usepackage[export]{adjustbox}
\defcitealias{Casado+15}{Paper~I}

\title
[Ageing and quenching on resolved scales.]
{Galaxy evolution on resolved scales: ageing and quenching in CALIFA.}

\author[Corcho-Caballero et al.]
{
P. Corcho-Caballero$^{1,2}$\thanks{E-mail:pablo.corcho@uam.es}, J.~Casado$^1$, Y.~Ascasibar$^{1}$ and R. Garc\'{i}a-Benito$^{3}$\\
$^{1}$Departamento de Física Teórica, Universidad Autónoma de Madrid (UAM), Campus de Cantoblanco, Madrid 28049, Spain\\
$^{2}$Australian Astronomical Optics, Macquarie University, 105 Delhi Rd, North Ryde, NSW 2113, Australia\\
$^{3}$Instituto de Astrof\'isica de Andalu\'ia (CSIC), E-18008, Granada, Spain
}

\date{\bf Draft version 4.0 (\today)}

\usepackage{color}

\newcommand{\change}[1]{#1}

\newcommand{\envir}{\ensuremath{R_5/\bar{r}}}
\newcommand{\ha}{\ensuremath{\rm H\alpha}}
\newcommand{\hb}{H$\beta$}
\newcommand{\oiii}{[OIII]}
\newcommand{\nii}{[NII]}

\newcommand{\shifu}{{\sc shifu}}
\newcommand{\be}{$f(\rm BE)$}
\newcommand{\ba}{$f(\rm BA)$}

\newcommand{\MgFe}{$\rm [MgFe]^\prime$}
\newcommand{\pp}{\citetalias{Casado+15}}

\usepackage{soul}

\begin{document}

\maketitle

\label{firstpage}

\begin{abstract}
This work investigates the fundamental mechanism(s) that drive galaxy evolution in the Local Universe. 
By comparing two proxies of star-formation sensitive to different timescales, such as EW(\ha) and colours like $g-r$, one may distinguish between smooth secular evolution (\emph{ageing}) and sudden changes (\emph{quenching}) on the recent star formation history of galaxies.
Building upon the results obtained from a former study based on ~80.000 SDSS single-fibre measurements, we now focus on spatially-resolved (on kpc scales) galaxies, comparing with a sample of 637 nearby objects observed by the CALIFA survey.
In general, galaxies cannot be characterised in terms of a single `evolutionary stage'.
Individual regions within galaxies arrange along a relatively narrow \emph{ageing sequence}, with some intrinsic scatter possibly due to their different evolutionary paths.
These sequences, though, differ from one galaxy to another, although they are broadly consistent with the overall distribution found for the (central) SDSS spectra.
We find evidence of recent quenching episodes (relatively blue colours and strong \ha~absorption) in a small fraction of galaxies (most notably, low-mass ellipticals), on global scales and individual regions (particularly at high metallicity).
However, we argue that most of the systems, over their entire extent, are compatible with a secular inside-out scenario, where the evolutionary stage correlates with both global (mass, morphology, and environment) as well as local (surface brightness and metallicity) properties.
\end{abstract}


\section{Introduction}
\label{sec:intro}

In order to constrain the physical mechanisms that regulate star formation and the relative importance of secular and environmental processes, we analysed in \citet[hereafter \pp]{Casado+15} the specific star formation rate (sSFR) for a sample of $\sim82500$ nearby galaxies extracted from the Sloan Digital Sky Survey DR7 \citep{DR7}.
By studying the dependence of the sSFR with other galactic properties, such as mass, metallicity, morphology, or neighbour overdensity, it was concluded that, although nurture processes do play a role and are indeed capable of suffocating star formation in some systems \citep[e.g.][]{Peng+2010, Peng+2015, Davies+2015, Davies+2019}, they are not dominant in the general case, and, \emph{on average}, it is the secular conversion of gas into stars that drives the evolution of galaxies \citep[e.g.][]{Abramson+15, Eales+18, CorchoCaballero+20}.
Following the terminology introduced in \pp, we will refer to this smooth secular evolution (over the whole life of the galaxy) as `ageing', and use the word `quenching' to denote a sudden truncation of the star formation activity due to a discrete event that takes place at a precise time.
Ageing is thus intrinsic to every object, even to those that suffer from violent nurture processes, whereas quenching may happen for an undetermined fraction of galaxies.

\begin{figure}

\centering
\includegraphics[width=.5\textwidth]{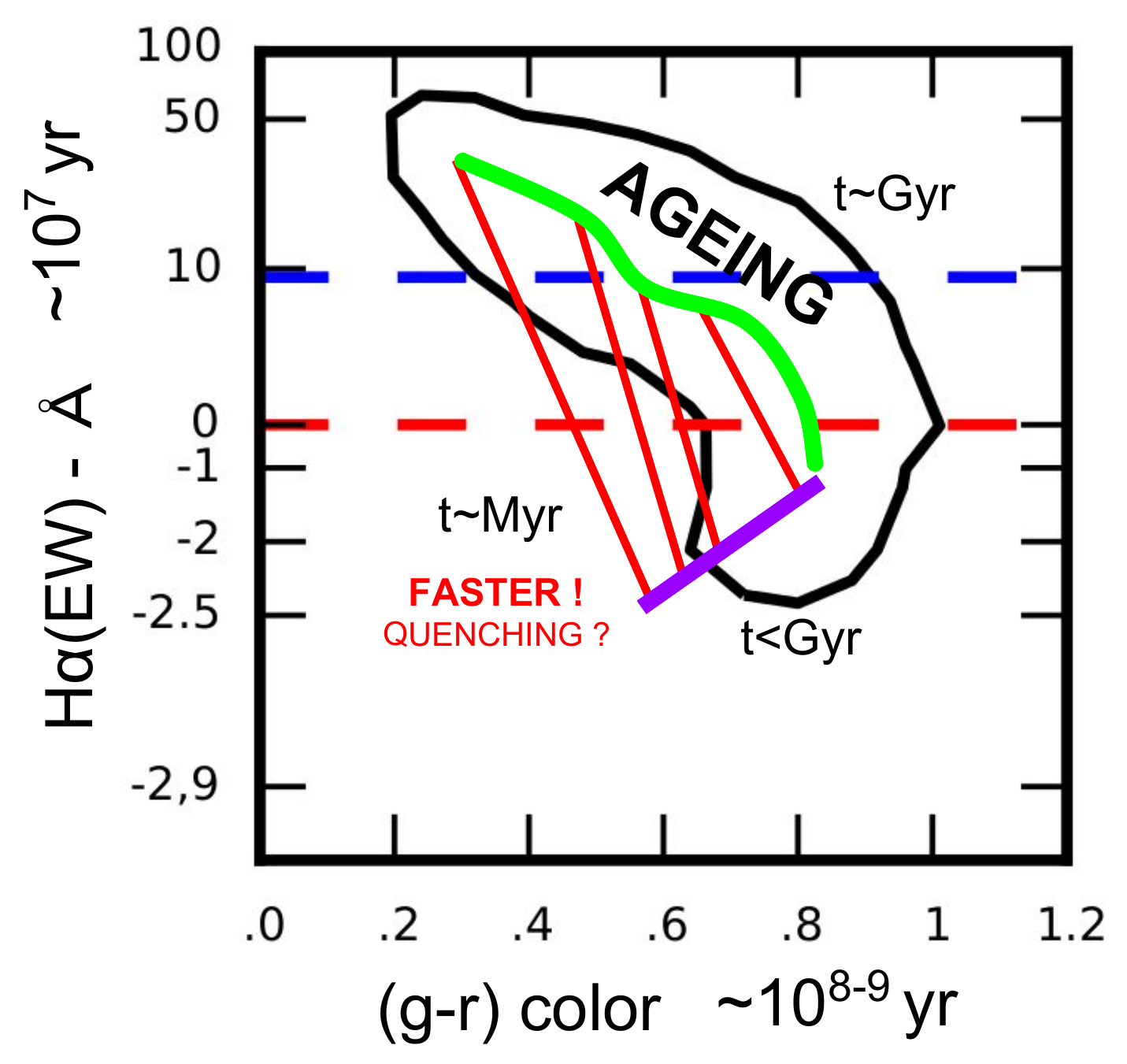}
\caption
{
Interpretation cartoon of the `ageing' of galaxies through the colour-equivalent width diagram (\pp). 
The black contour corresponds to the 90\% of the sample considered in \pp\ (SDSS). 
Blue dashed line \ha(EW)$=9$ is roughly equivalent to the EW threshold imposed in the diagram if we consider S/N$>$3 in all 4 BPT lines to derive a star forming sample (see \pp). Red dashed line corresponds to the zero value of EW. The equivalent width of \ha\ (y-axis) traces star formation on timescales of the order of $10^{7}$yr, as its emission is dominated by the light of O and B stars. 
The (g-r) colour (x-axis) traces SF on timescales of the order of $10^{8.5}$yr, as it is dominated by the emission of the continuum and evolves more slowly and smoothly due to the broader wavelength range covered. Green solid line corresponds to the `ageing sequence', our interpretation of a slow, smooth, secular evolutionary mode (nature) in galaxies directly related to the consumption (and conversion) of gas into stars. 
Steeper red lines represent a much abrupt (faster) transition through the diagram and is an observed trend that we aim to relate to nurture processes such as strangulation, quenching, etc. 
Purple solid line corresponds to our posed `quenching sequence'.
}
\label{fig_ageing_schema}
\end{figure}

As proposed in \pp, the effects of ageing and quenching can be investigated by means of the colour-equivalent width diagram, whose physical interpretation is qualitatively illustrated in Figure~\ref{fig_ageing_schema}.
Both quantities are proxies for the sSFR, but they are sensitive to different stellar populations and therefore to different time scales.
On the one hand, the intensity of the \ha\ line is proportional to the number of O and B stars, responsible for the bulk of ionizing ultraviolet radiation.
When considered in terms of equivalent width (EW), it roughly traces the fraction of stellar mass that has formed over the last $10-30$~Myr.
On the other hand, broad-band colours are more sensitive to the overall shape of the spectral energy distributions, and they trace older stellar populations, varying on time scales of the order of $\sim 0.1-1$~Gyr.

If the star formation rate varies smoothly over the whole life of a galaxy, the system slowly moves from the `primitive' extreme on the upper left -- blue colours and high EW, which also imply low metallicity and high gas fraction \citep[see e.g.][]{Ascasibar+15, Casado+15} -- towards the `evolved' extreme on the bottom right (red colours and EW of the order of $1-2$~\AA\ in \emph{absorption}, which we denote as negative values).
One of the main results of \pp\ is indeed that most galaxies in the field describe a relatively tight `ageing sequence' in the colour-equivalent width plane (\change{green line} in Figure~\ref{fig_ageing_schema}), akin to the `spectroscopic sequence' proposed in \citet{Ascasibar+11}, and such arrangement was interpreted in terms of gradual evolution driven by the secular conversion of gas into stars.
At the present time, some galaxies are more `chemically primitive' and some others are more `chemically evolved'.
There are strong correlations with mass and morphology, but we found no evidence of a bimodal population of `star-forming' and `passive' systems; only different levels of star formation activity, and it was argued that the `main sequence' of galaxy formation \citep[approximately constant sSFR for `star-forming' galaxies; see e.g.][and references therein]{Noeske+07, Speagle+14, Cano-Diaz+16, Oemler+17, Hsieh+17, Cano-Diaz+19, Davies+19a, Caplar&Tacchella19} may be largely due to a selection effect, associated to an arbitrary distinction between both populations \citep[e.g.][]{Eales+18, CorchoCaballero+20}.

In the densest environments, spiral galaxies are also found along a similar ageing sequence, although they tend to be more chemically evolved than objects in the field (for a given stellar mass, they have systematically lower sSFR and older stellar populations).
However, their star formation rate does not show any obvious signature that it has undergone any sudden change in the recent past.
In contrast, many early-type galaxies in dense environments are arranged in a tight disposition with stronger \ha\ absorption in bluer systems, consistent with the quenching scenario.
If star formation is suddenly quenched, the intensity of the \ha\ emission line drops on a time scale of the order of $10-30$~Myr (the lifetime of O and B stars), but the system will display intermediate colours and very strong absorption lines due to the presence of A stars.
As these evolve and die (over the next $\sim 300$~Myr), the galaxy moves towards less negative EW and redder colours, until they finally join the evolved end of the ageing sequence, along a track in the colour-equivalent width diagram (purple solid line in Figure~\ref{fig_ageing_schema}) that was dubbed `quenched sequence' in \pp.
\begin{figure*}
\centering
\includegraphics[width=\textwidth]{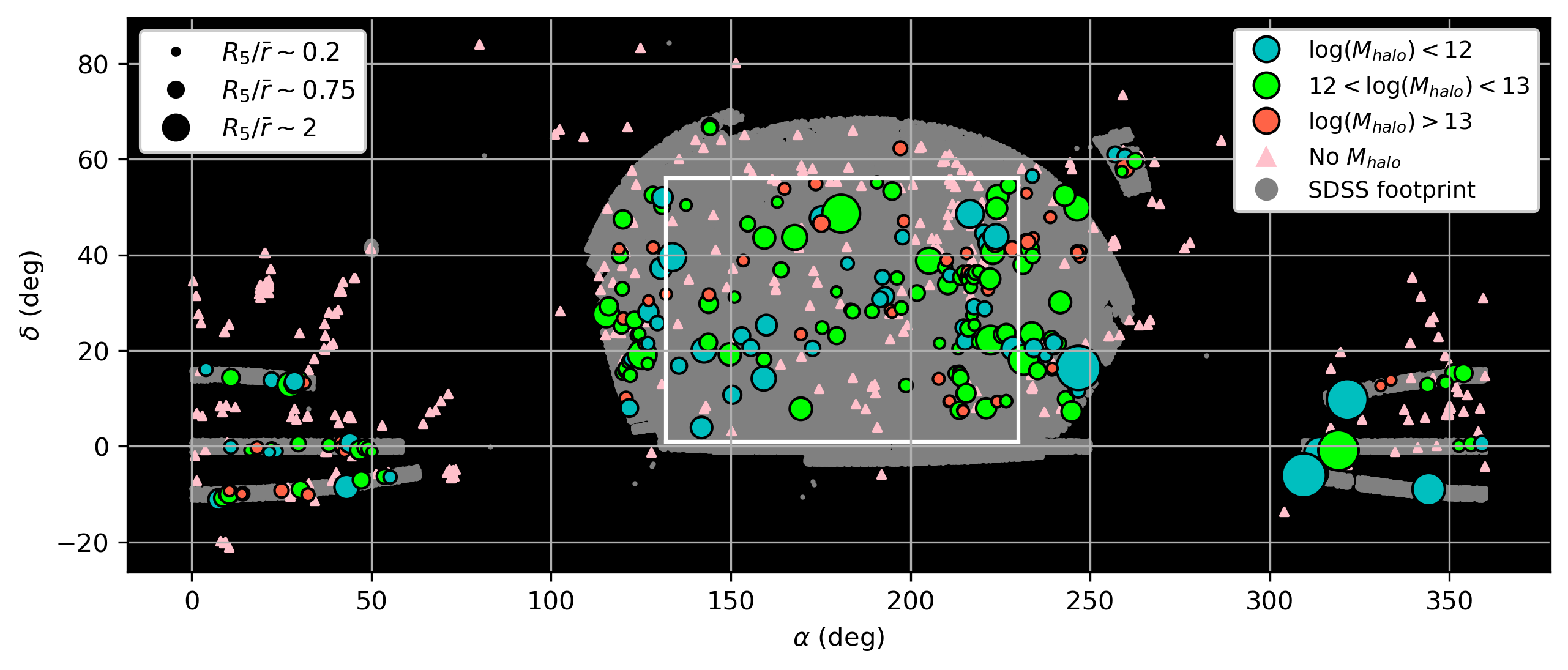}
\caption
{
CALIFA selected sample over SDSS spectroscopic footprint.
Galaxies where both environment proxies ($R_5/\bar{r}$, $M_{vir}$) are available are denoted by colored circles, with sizes indicating the normalized distance to the fifth nearest neighbour and colors indicating the halo group mass (see legend).
White square denotes the SDSS sample region used in \pp.}

\label{fig:sample_footprint}
\end{figure*}

However, all these results, based on SDSS measurements of very nearby galaxies, may be severely affected by aperture bias \citep[e.g.][]{DuartePuertas+17, Sanchez+20}.
The $3''$ fibre translates into a physical diameter of $1.2-4$~kpc for the redshift range considered in \pp\ ($0.02<z<0.07$), and it only covers the brightest part (typically the centre) of the objects under study.
In the present work, we extend the study of the colour-equivalent width diagram to a sample of 637 objects observed by the Calar Alto Legacy Integral-Field Area (CALIFA) survey \citep{Sanchez+12}.
The field-of-view (FoV) of the PPAK/PMAS instrument consists of an hexagon of $65''\times 74''$, and the CALIFA sample has been diameter-selected so that it typically reaches up to about $\sim 2$ effective radii, ensuring a representative coverage of the observed galaxies \citep{Walcher+14}.
With these integral-field spectroscopic (IFS) data \citep[see][for a review]{Sanchez20ARA&A}, we will investigate the ageing and quenching processes on resolved ($\sim$kpc-size) scales over the whole galactic extent.


The galaxy sample considered for this work and its physical characterization are briefly described in Section~\ref{sec:observations}. We explore the colour-EW diagrams for every given object and the results of our analysis are presented in Section~\ref{sec:results} and discussed in Section~\ref{sec:discussion}. 
Finally we present a summary and our conclusions in Section~\ref{sec:conclusions}.

\section{Observational data}
\label{sec:observations}
All data used in \pp\ correspond to objects catalogued as galaxies by the spectroscopic pipeline (table SpecObj) in the SDSS Data Release 7 database\footnote{\url{http://cas.sdss.org/dr7/en/help/browser/browser.asp}} \citep{DR7}.
The sample is restricted in apparent magnitude $m_{\rm r}<17.5$ to ensure that every object with a similar-mass companion is properly identified, and limited to a region of the sky (white square in Figure~\ref{fig:sample_footprint}) to minimize boundary effects in the neighbour search.
We also imposed a redshift range $0.02 < z < 0.07$, where the upper limit implies completeness for absolute magnitude $M_r<-20.0$, and the lower redshift cut avoids local structure.
The final sample consisted of $\sim 82500$ objects, for which several galactic properties were derived (see \pp\ for a detailed description).

The main purpose of the present work is to study in more detail the colour-EW diagram using IFS observations.
To do so we have drawn a sample of 637 nearby galaxies ($0.002<z<0.08$) observed within the CALIFA collaboration\footnote{\url{http://califa.caha.es/}} from the mother (529) and extension (111) samples, in the V500 (low-resolution) setup from the $3^{\text{rd}}$ Data Release \citep{dr3}.
Galaxies in the mother sample have well defined selection criteria \citep{Walcher+14}, ensuring both a reasonably homogeneous spatial sampling and an optical extension in the $r$-band up to 2 effective radii.
On the other hand, galaxies belonging to the extension sample provide a wealth of rarer objects complementary to the mother sample: dwarf systems (28), interacting galaxies (27), low- and high-mass early type galaxies (31), low-mass galaxies that hosted Type Ib, Ic and II supernovae (14) or compact early-type galaxies hosting supermassive black holes at their centres (7).
Therefore, the joint sample is expected to comprise a large variety of evolutionary pathways, which may in principle display different trends in the color-EW diagram.

\begin{figure}
\centering
\includegraphics[width=\linewidth]{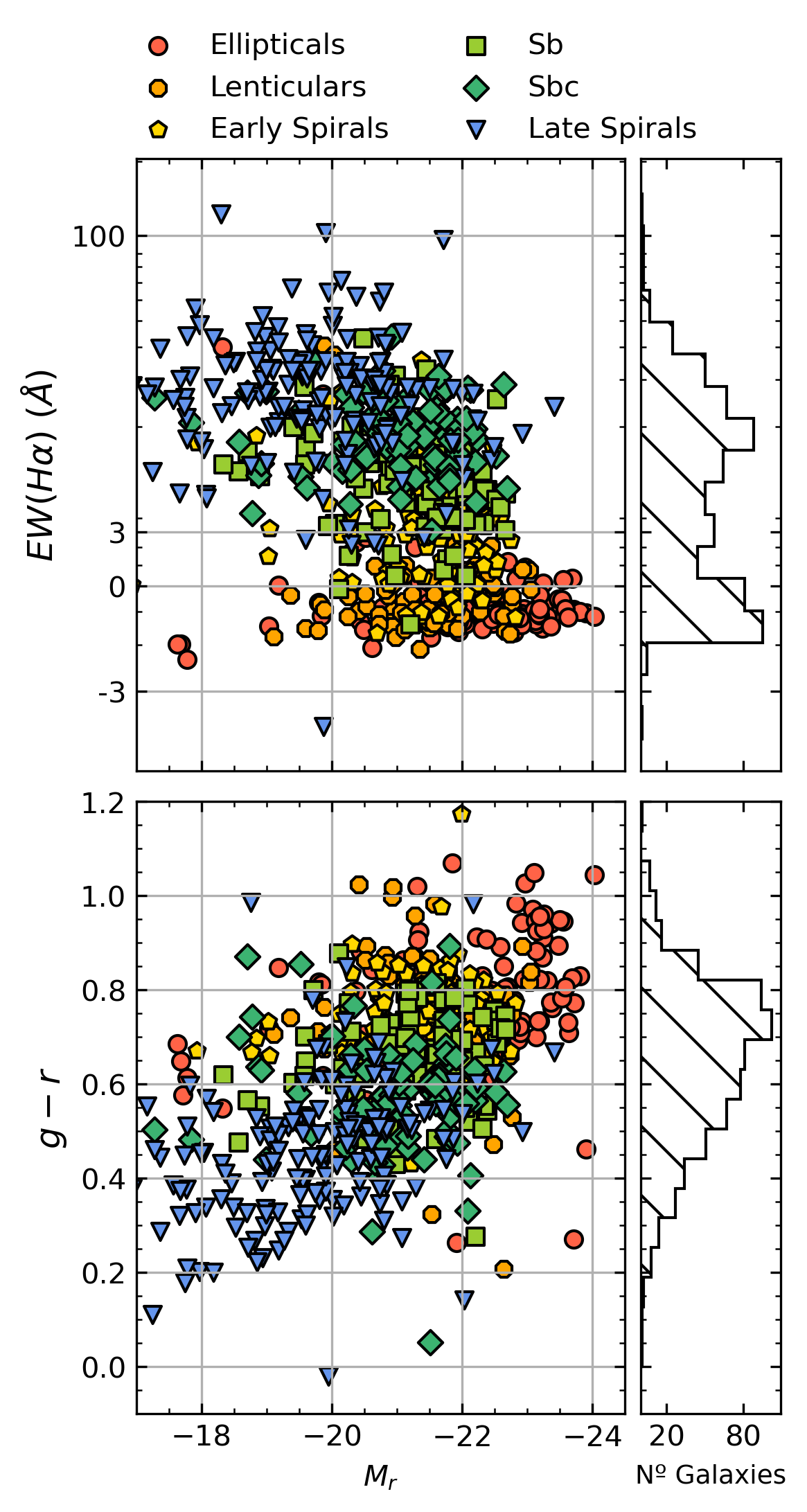}
\caption
{
Location of the sample in the EW(\ha)-magnitude (top) and color-magnitude (bottom) diagrams.
The colour code refers to the morphological bins we define in Section~\ref{sec:global_properties} following the \citet{Walcher+14} classification.
}
\label{sample_global_SDSS}
\end{figure}
\begin{figure}
\centering
\includegraphics[width=\linewidth]{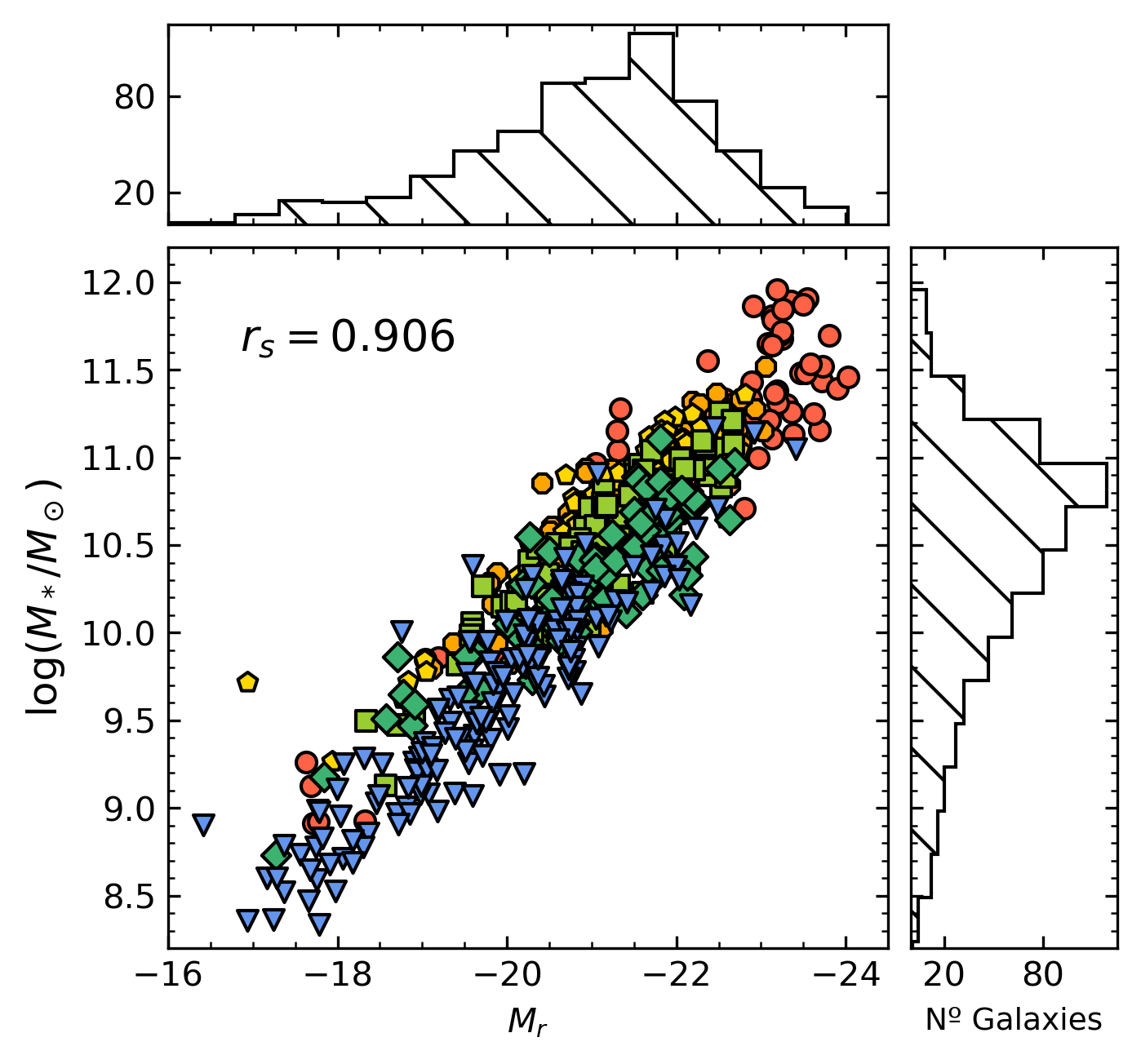}
\caption
{
Plot comparing the total stellar masses estimated using growth curve magnitudes from SDSS, GALEX and 2MASS 
\citep[$M_{*}$, ][]{Walcher+08} with the \emph{model} absolute magnitude in the r band (M$_{r}$, proxy used in \pp) for the entire sample. 
The colour coding is the same as in Figure~\ref{sample_global_SDSS}.
Spearman's rank correlation coefficient, $r_s = 0.906$, is quoted in the top left corner.}

\label{fig_mass}
\end{figure}

For illustrative purposes, the CALIFA sample used in the present work is represented in Figure~\ref{fig:sample_footprint} together with the SDSS spectroscopic catalog as grey points.
Galaxies for which group mass measurements are available (see sec~\ref{sec:global_properties} for further details on environmental estimators) are denoted by circles indicating the normalised distance to their fifth nearest neighbour (size) and group mass (color).
Otherwise they are denoted as pink triangles. 
In spite of the large difference regarding the number of galaxies used in \pp\ and the present work, it must be noted that the total number of \emph{spectra} analysed is now much larger.
Since each datacube contains of the order of $2000-2500$ useful spaxels, our CALIFA sample comprises nearly $\gtrsim 1$ million individual spectra.

The main advantage of using IFS data relies on the possibility to study the spatial distribution of star-forming/quenching processes over a significant fraction of the total extent of galaxies.
Rather than treating galaxies as single points that define an ageing sequence, as in \pp, we will now focus on the evolutionary state of galaxies individually, in terms of their resolved distribution over the \emph{ageing diagram}.



\subsection{Physical characterization}
\label{sec:physical_characterization}

As we aim to reproduce, validate and extend the analysis conducted in \pp\ we will again briefly outline how the sample was characterized in the previous work. 
We used the absolute magnitude in the $r$ band, $M_r$ as a proxy for the stellar mass. 
To quantify the environment in terms of the local galaxy overdensity we derived the projected distance to the fifth nearest neighbour normalized to the average intergalactic distance $\overline{r}$(z).
We also made use of the visual morphology classification conducted by the Galaxy Zoo 1 citizen science project \citep{Lintott+08}. 
Finally, the (u-r) colour (SDSS filters) and the equivalent width of the \ha\ line were used as proxies of the specific star formation rate. 

In the present work we will use some of these values to characterize the global properties of the CALIFA sample.
Besides, the environmental characterization, \envir, was derived making use of the entire SDSS spectroscopic sample and was proven to be a robust indicator of the local galaxy overdensity. 
The availability of integral-field spectroscopy data will allow us to measure galaxy colours ($g-r$) and the EW(\ha) with a better spatial sampling and size coverage.


Figure~\ref{sample_global_SDSS} shows the location of the CALIFA sample in the EW(\ha)-magnitude (top) and colour-magnitude (bottom) diagrams according to their integrated values and their corresponding distributions (top- and bottom-right histograms, respectively).
Throughout this paper, unless explicitly noted otherwise, we will define equivalent width as
\begin{equation}
\label{eq:ew}
{\rm EW(H\alpha)}
\equiv
\int_{\rm 6550\,\AA}^{\rm 6575\,\AA}
  \frac{ F_\lambda(\lambda) }
  {\frac{F_{\rm B}\lambda_{\rm R}-F_{\rm R}\lambda_{\rm B}}{\lambda_{\rm R}-\lambda_{\rm B}}+
   \lambda\frac{F_{\rm R}-F_{\rm B}}{\lambda_{\rm R}-\lambda_{\rm B}}}-1\ {\rm d}\lambda
\end{equation}
where $F_{\rm B}$ and $F_{\rm R}$ correspond to the mean flux per unit wavelength computed in the galaxy rest-frame $6470-6530$~\AA\ and $6600-6660$~\AA\ bands, with central wavelengths $\lambda_{\rm B}=6500$~\AA\ and $\lambda_{\rm R}=6630$~\AA, respectively. 
Under this definition, positive and negative values of EW denote emission and absorption, respectively.

In terms of EW(\ha), we can see that the present sample covers homogeneously a wide range of values: from late spirals (blue triangles) with high \ha~emission to ellipticals (red circles) with \ha~absorption features. 
Galaxy colours are also broadly distributed, with most systems presenting intermediate values around $(g-r)\sim0.7$, typical of the Green Valley \citep[e.g.][]{Blanton+03}.
Therefore, this sample provides a good opportunity to explore, by means of the color-EW diagram, the possible \emph{quenching} or \emph{ageing} mechanisms that operate in the Local Universe in order to drive galaxies through the Green Valley towards the red sequence \citep[e.g.][]{Schawinski+14}.


\begin{figure}
\centering
\includegraphics[width=\linewidth]{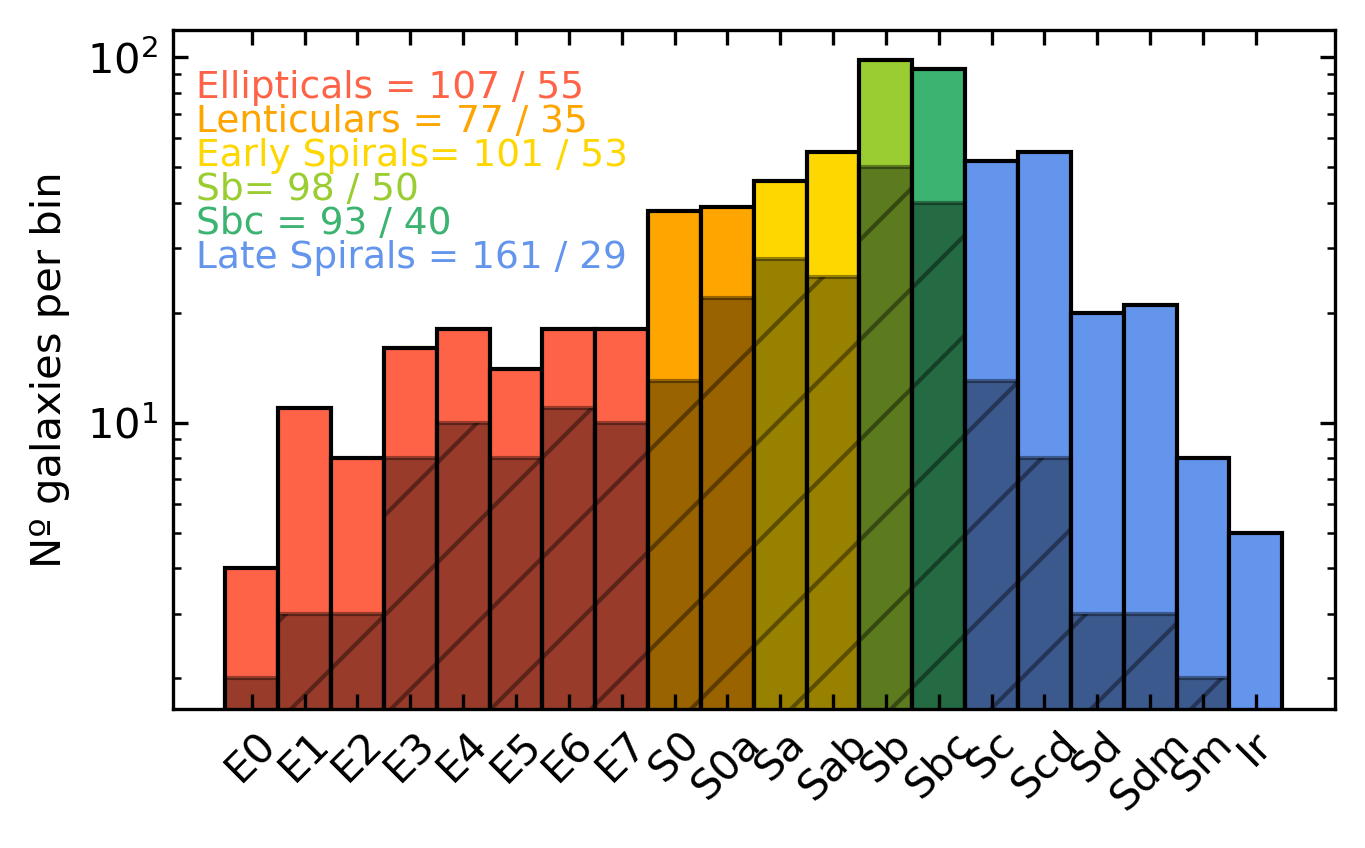}
\caption
{
Visual morphological classification of the 637 objects in the CALIFA sample, according to \citet{Walcher+14}. Coloured histogram denotes the complete sample while black-hatched histogram represents galaxies with environmental information available. 
The number of galaxies in each bin is quoted for both histograms on the top-left corner of the figure.}
\label{fig:sample_morpho}
\end{figure}
\subsubsection{Global properties}
\label{sec:global_properties}

In \pp\ we used the absolute magnitude in the $r$ band, M$_{r}$, as a proxy for stellar mass.
However, the CALIFA collaboration has a stellar mass catalogue for its entire dataset computed from the SDSS optical $ugriz$, GALEX UV, and 2MASS NIR growth curve magnitudes using the prescription presented in \citet{Walcher+08}.
We consider this value in our analysis, which is shown in Figure~\ref{fig_mass} to be in good agreement with the proxy used in \pp. 
As noted above, the present sample provides a wide mass coverage from dwarf systems with $\log(M/M_\odot)\lesssim 9$ to extremely massive galaxies with $\log(M/M_\odot)\gtrsim 11.5$, although the vast majority of objects lies in the Milky-Way regime $10 \lesssim \log(M/M_\odot)\lesssim 11$.

At a given fixed mass, morphology is one of the global properties of a galaxy that is most strongly correlated with its star formation history.
In \pp\ we used Galaxy Zoo morphological classification, according to which a large number the galaxies were assigned to the `unknown' category.
Given the correlation between SF and morphology, these `unknown' objects are believed to represent a transition population, the most interesting objects for exploring ageing and quenching processes.
In the present work, we make use of the morphological classification described in \citet{Walcher+14}, based on visual inspection of SDSS images in the $r$ and $i$ bands.
We further bin our sample into 6 different groups based on \citet{Walcher+14} morphologies: ellipticals (E0-E7), lenticulars (S0-S0a), early-spirals (Sa-Sab), Sb's, Sbc's and late-spirals (Sc-Sm+Irregulars).
The number of galaxies in each mophological group is summarised in Figure~\ref{fig:sample_morpho}

\begin{figure}
\centering
\includegraphics[width=.47\textwidth]{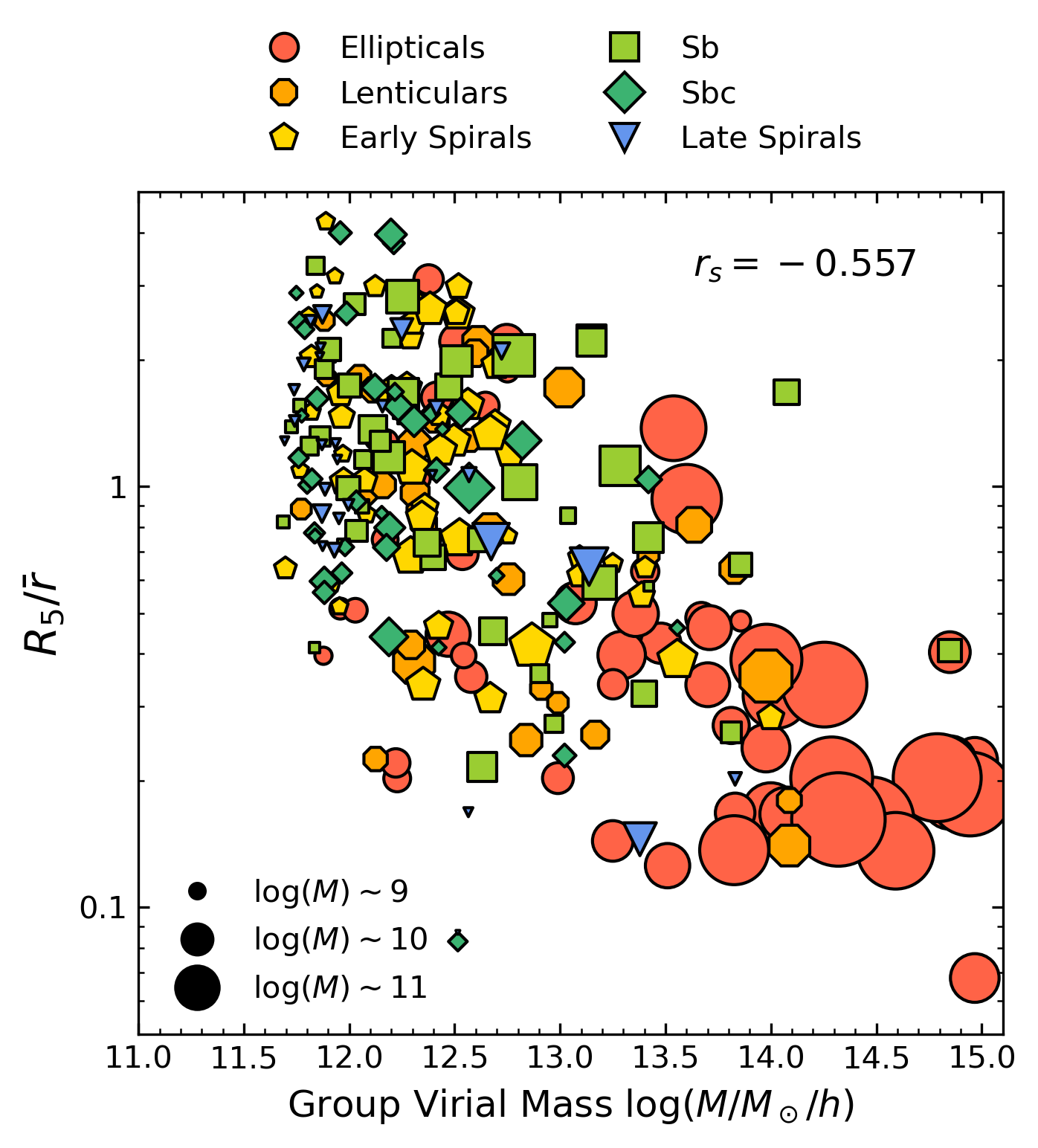}
\caption
{
Comparison between the group virial mass $M_{\rm vir}$ derived by \citet{Yang+07} and the overdensity \envir\ measured in \pp\ for our CALIFA sample.
Colour coding is the same as in Figure~\ref{sample_global_SDSS}, and symbol size indicates stellar mass (see legend).
Spearman's rank correlation coefficient, $r_s -= 0.557$, is quoted in the top right corner.
}
\label{fig:environment_sample}
\end{figure}
\begin{table*}
\begin{center}
\setlength{\tabcolsep}{3pt} 
\begin{tabular}{c|c|c|c|c|c|c|c|c|c|c|c|c|c}
\hline
$\!\!\!\!$CALIFAID$\!\!\!\!$ & NAME & Morpho & \envir & $\log(\frac{M_{\rm vir}}{M_\odot})$ & $\log(\frac{M_*}{M_\odot})$ & M$_{r}$ &  $\log(\rm\frac{[NII]}{H\alpha})$ & $\log(\rm\frac{[OIII]}{H\beta})$ & $\rm \frac{EW_{em}(H\alpha)}{\AA}$ & $f(\rm{BE})$ & $f(\rm{BA})$ & $f(\rm{RE})$ & $f(\rm{RA})$ \\
\hline
1 & IC5376 & Sb & n/a & n/a & 10.54 & -21.12 & -0.093 & 0.201 & 1.288 & 0.392 & 0.002 & 0.584 & 0.022 \\ 
2 & UGC00005 & Sbc & n/a & n/a & 10.73 & -22.16 & 0.005 & 0.579 & 6.17 & 0.911 & 0.000 & 0.089 & 0.000 \\ 
3 & NGC7819 & Sc & n/a & n/a & 10.1 & -21.16 & -0.454 & -0.729 & 33.282 & 0.993 & 0.000 & 0.007 & 0.000 \\ 
5 & IC1528 & Sbc & n/a & n/a & 10.23 & -21.04 & -0.43 & -0.64 & 17.007 & 0.981 & 0.000 & 0.019 & 0.000 \\ 
6 & NGC7824 & Sab & n/a & n/a & 11.09 & -22.33 & n/a & n/a & n/a & 0.281 & 0.184 & 0.282 & 0.253 \\ 
7 & UGC00036 & Sab & n/a & n/a & 10.9 & -21.73 & -0.208 & -0.165 & 2.835 & 0.451 & 0.006 & 0.541 & 0.003 \\ 
8 & NGC0001 & Sbc & n/a & n/a & 10.58 & -21.75 & -0.385 & -0.471 & 6.148 & 0.740 & 0.000 & 0.260 & 0.000 \\ 
9 & NGC0023 & Sb & n/a & n/a & 10.83 & -22.53 & -0.265 & -0.432 & 23.113 & 0.887 & 0.001 & 0.112 & 0.000 \\ 
10 & NGC0036 & Sb & n/a & n/a & 10.9 & -22.38 & 0.201 & 0.505 & 0.524 & 0.580 & 0.003 & 0.288 & 0.129 \\ 
... & ... & ... & ... & ... & ... & ... & ... & ... & ... & ... & ... & ... & ... \\
\hline
\end{tabular}
\end{center}
\caption
{
Global properties of the galaxies in the CALIFA sample.
`CALIFAID' corresponds to the label assigned by the collaboration, while `NAME' corresponds to the NGC, UGC, or NED catalogue.
`Morpho' refers to the visual morphological classification conducted by \citet{Walcher+14};
`\envir' is the environmental characterization performed in \pp, and `$M_{\rm vir}$' corresponds to the group virial mass in \citet{Yang+07}.
$M_{*}$ corresponds to the stellar mass derived by \citet{Walcher+08}; M$_{r}$ is the \emph{modelmag} magnitude in the $r$ band obtained from the SDSS DR7 database; $\log(\rm\frac{[NII]}{H\alpha})$, $\log(\rm\frac{[OIII]}{H\beta})$ and $\rm EW_{em}(H\alpha)$ correspond to the values derived to perform the nuclear classification as discussed in \ref{sec:global_properties};
$f(\rm{BE})$, $f(\rm{BA})$, $f(\rm{RE})$ and $f(\rm{RA})$ denote the fraction of regions in each of the four domains (blue-emission, blue-absorption, red-emission, and red-absorption) of the ageing diagram as discussed in \ref{sec:results}.
Only the top 10 entries are shown; the complete table is available in the online version of the journal.
}
\label{tab_fsample}
\end{table*}

Additionally, we have used \envir, the normalized distance to the fifth `spectroscopic' neighbour, as a proxy to characterize the environment (see \pp\ for a detailed description).
In order to avoid boundary effects, we have computed \envir~only for those galaxies with at least one neighbour in the redshift range $z_i\pm0.005$ within a sphere of 1 degree radius.
We will compare our proxy, \envir, with the estimation for the group virial mass computed from the SDSS group catalogue created by \citet{Yang+07}.
These authors used a group finder to determine group membership and halo masses within the SDSS DR7 sample of galaxies.
Their algorithm iteratively identifies potential group members, determines a characteristic luminosity for the defined group, and estimates the mass, size and velocity dispersions of the dark matter halo.
To do so, it assumes a group mass-to-light ratio adjusted at every iteration, and the group membership is updated until convergence is reached.
As shown in Figure~\ref{fig:sample_footprint}, only 263 galaxies in our CALIFA sample have environmental proxies (\envir, $M_{\text{vir}}$) available.
This value is listed as `group virial mass' in Table~\ref{tab_fsample} together with \envir.
Figure~\ref{fig:environment_sample} shows that they are strongly correlated: CALIFA galaxies in dense environments typically show $M_{\rm vir}\gtrsim 13$ and \envir$\lesssim0.5$, with a predominant fraction of massive systems with early-type morphologies.
On the other hand, \emph{field} galaxies are in general found in less massive halos ($M_{\rm vir}\lesssim 12.5$) and isolated regions (\envir$\gtrsim0.7$), mainly composed by a mixture of low-mass objects with disk-like morphologies.

Finally, we will also consider as a global property the nuclear classification according to the BPT diagram \citep{BPT} for the subsample of 433 galaxies where measurements of the intensity of the \ha, \hb, \oiii\ and \nii~emission lines within the central kpc region are possible from the CALIFA COMBO datacubes.
To account for stellar absorption, we use {\sc starlight} \citep{Cid-Fernandes+05} to model the underlying stellar population as a linear combination of single stellar populations spanning different ages and metallicities.
We use the spectra provided by \citet{Vazdekis+10} for populations older that 64~Myr and \citet{Gonzalez-Delgado+05} models for younger ages as in \cite{deAmorim+2017}.
Dust effects are modelled as a foreground screen, assuming a \citet{Cardelli+89} reddening law with $R_V = 3.1$.
Then, we subtract the estimated stellar continuum from the observed spectrum and obtain the line flux with the SHerpa IFU line fitting software (\shifu; Garc\'{i}a-Benito, in preparation), based on CIAO’s {\sc Sherpa} package \citep{Freeman+01, Doe+07}.
Small deviations with respect to the stellar continuum are taken into account by a first order polynomial, and independent Gaussians have been fitted for the emission lines.
Galaxies above the \citet[]{Kewley+01} demarcation line will be simply interpreted as \emph{AGN/Shock}, whereas galaxies below the \citet[]{Kauffmann+03} demarcation line will be labeled as \emph{star forming}. 
Intermediate values between both regimes will be classified as \emph{composite}.
The nebular \emph{emission} equivalent width EW$\rm _{em}(H\alpha)$, after accounting for stellar absorption, has been argued to provide a complementary discriminator between different ionising conditions, with strong AGN and high-velocity shocks presenting EW$\rm _{em}(H\alpha)$ $\gtrsim 6$~\AA, whereas Hot Low-Mass Evolved Stars (HOLMES) and post-AGBs feature values EW$\rm _{em}(H\alpha)$ $\lesssim3$~\AA.
Galactic nuclei with $3 \lesssim$ EW$\rm _{em}(H\alpha)$ $\lesssim 6$~\AA\ are classified as weak AGN \citep[see][for a detailed discussion]{Sanchez+21}.

The global properties of our galaxies are provided in Table~\ref{tab_fsample}.

\subsubsection{Local properties}
\label{sec:rgb}

Accurately measuring low values of EW(\ha) on resolved scales poses a challenge for current IFS data. 
Typically, those spaxels correspond to regions with low signal to noise ratio (S/N), especially at the outer parts of galaxies.
In order to mitigate this issue we have applied a Voronoi binning  \citep{Cappellari&Copin03} with target S/N$=30$ to each datacube in our sample, based on flux maps computed in the spectral range 6540-6580 \AA.
Under this constraint, each datacube effectively comprises a few hundred independent regions.
We have masked foreground stars as illustrated on the individual images presented in Appendix~\ref{appendix:individual_distributions}. 

We will consider the surface brightness in the $r$ band $\mu_r$ and the $[\rm MgFe]^\prime=\sqrt{\text{Mgb}(0.72~Fe5270+0.28~Fe5335)}$ index defined by \citet{Thomas+03} as observational proxies for stellar mass surface density and stellar metallicity, respectively.

In general, high values of the surface brightness can only be reached in the innermost part, close to the galactic centre, whereas low values are representative of the outskirts.
As a rule of thumb, an SDSS $r$-band surface brightness of the order of 22~mag/arcsec$^2$ roughly corresponds to one effective radius, $R_e$.
On the other hand, gas-phase metallicity has been shown to correlate with stellar surface density \citep{Sanchez+13} and stellar-to-gas fraction \citep{Ascasibar+15} on local scales.
Physically, it seems reasonable to expect that higher surface brightness and metallicity are associated to shorter timescales and a more efficient conversion of gas into stars.
According to this argument, the central parts should be more `chemically evolved' than the outer regions, consistent with the inside-out galaxy formation scenario \citep{perez+13, gonzalez-delgado+14, IbarraMedel+16, garcia-benito+17}.

\section{Results}
\label{sec:results}

\begin{figure*}
    \centering
    \includegraphics[width=1\linewidth]{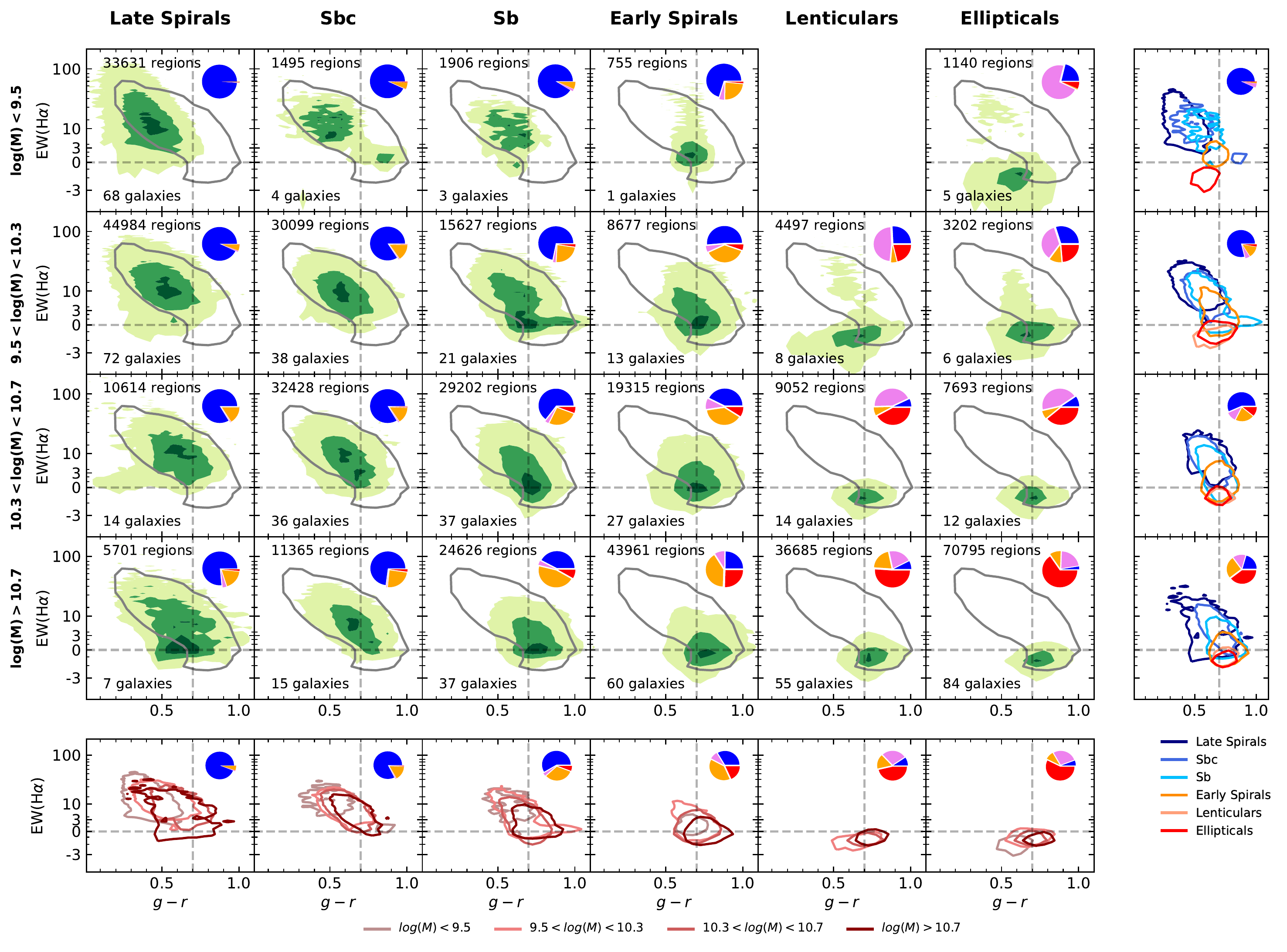}
    \caption{
Dependence of the color-EW diagram on global galaxy mass and morphology.
Green contours on the main panels encompass 90, 50 and 10 percent of the \change{total probability distribution corresponding to all galaxies within each bin.}
The number of galaxies and regions used is shown at the bottom and top of each panel, respectively.
Pie charts on the top-right corner illustrate the fraction of regions in the 4 domains (separated by dashed grey lines) of the ageing diagram: blue-emission (blue), blue-absorption (pink), red-emission (orange) and red-absorption (red).
A grey solid line indicates the contour containing 90\% of the galaxies obtained for the $\sim 82500$ SDSS galaxies ($3''$ fibres) in \pp. 
Complementary to the main panels, an additional row at the bottom compares the 50 percent contours of the different mass bins for a given morphological group.
Conversely, an additional column on the right displays different morphologies for each mass bin (see legends).
}
\label{fig:distrib_morph_mass}
\end{figure*}

Here we will explore the impact of the global and local properties defined above on the ageing diagram. 
For each galaxy, we select all Voronoi regions with surface brightness $\mu_r \leq 24~\text{mag/arcsec}^2$, whose uncertainties are assumed to follow a bivariate normal distribution with no correlation between EW(\ha) and $(g-r)$, $\mathcal{N}(g-r,~\mathrm{EW},~\sigma_{\mathrm{EW}}^{2},~\sigma_{g-r}^{2})$.

In order to provide a straightforward characterization of the ageing diagram, we have divided it into four different domains defined as
\begin{enumerate}
    \item blue-emission (BE): EW(\ha)$>0$ and $(g-r)<0.7$
    \item blue-absorption (BA): EW(\ha)$\leq0$ and $(g-r)<0.7$
    \item red-emission (RE): EW(\ha)$>0$ and $(g-r)\geq 0.7$
    \item red-absorption (RA): EW(\ha)$\leq0$ and $(g-r)\geq0.7$
\end{enumerate}
Galaxy regions located within BE are frequently star-forming, whereas those in the BA domain would be candidates to have experienced recent quenching episodes.
The RE domain can be interpreted as the old-stellar-population regime with residual star formation and/or \ha~emission from other ionizing sources. 
Finally, `retired' regions are found at RA with red colors and absorption features.
The ageing scenario can be roughly interpreted as a smooth evolutionary sequence $\rm BE\xrightarrow{\ga 1~Gyr} (RE) \xrightarrow{\ga 1~Gyr} RA$\footnote{\change{
Note that many galaxy regions may be located at RE simply due to the effect of dust extinction.}
} over the age of the universe, whereas we understand quenching as a rapid transition $\rm BE\xrightarrow{\sim 20~Myr} BA\xrightarrow{\sim 0.5~Gyr}RA$.
Table~\ref{tab_fsample} includes the fraction of regions that are located within each domain for all galaxies in our sample.

\subsection{The role of global properties}
\label{sec:results_global_properties}
\begin{figure*}
    \centering
    \includegraphics[width=0.9\linewidth]{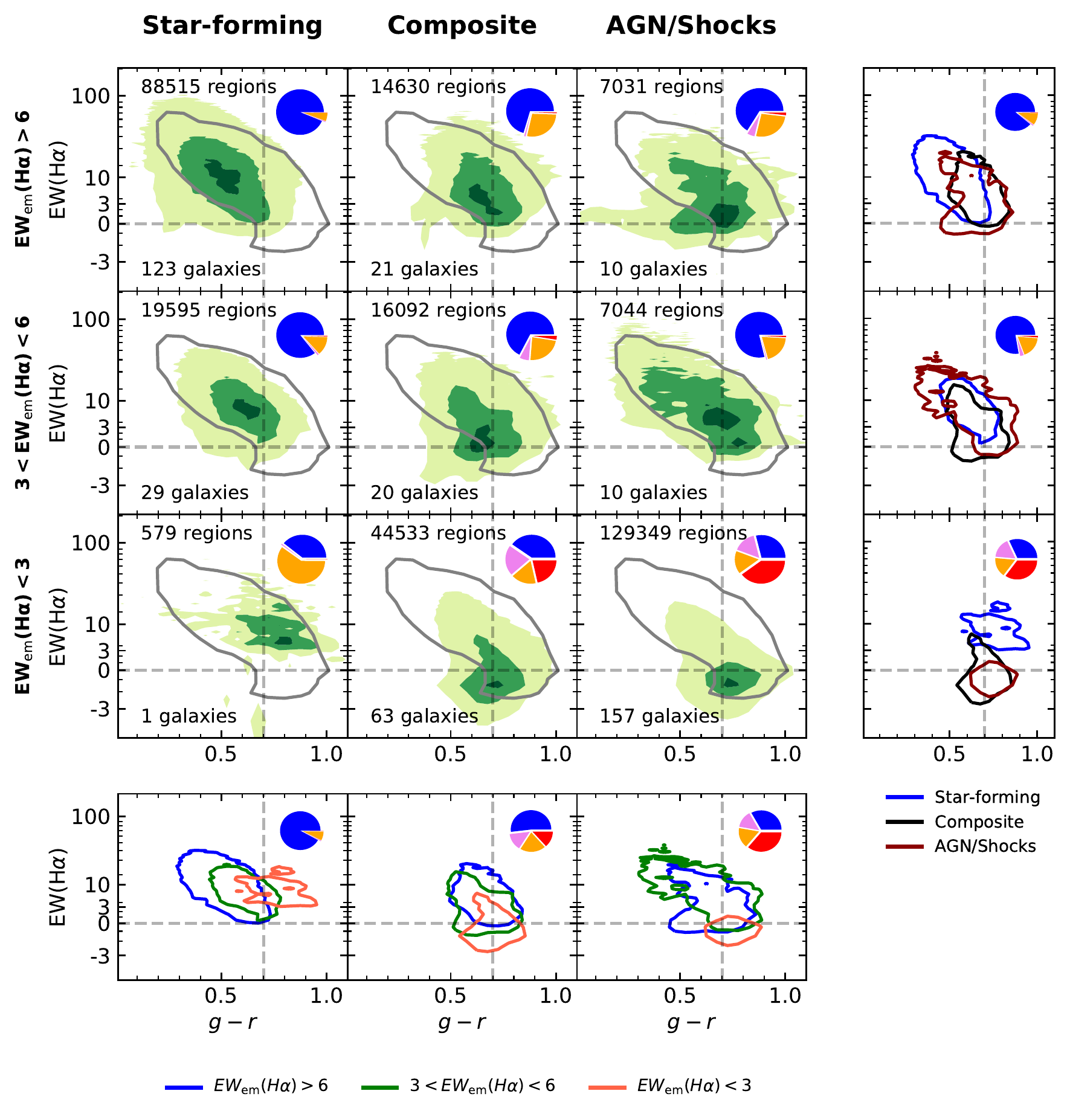}
    \caption{Dependence of the color-EW diagram on nuclear properties.
    This figure is arranged similarly to Fig.~\ref{fig:distrib_morph_mass} but binned in terms of EW(\ha)$_{\rm em}$ (rows), and BPT classification (columns). }
    \label{fig:distrib_bpt_ew}
\end{figure*}

\change{In this section galaxies will be binned in terms of their global stellar mass, morphology, nuclear activity, and environment.
For each galaxy $i$, the probability distribution in the ageing diagram is given by the normalised sum over all Voronoi regions $\nu$
\begin{equation}
\Phi_i(g-r, {\rm EW}) \equiv
    \frac{ {\rm d} f_i (g-r, {\rm EW}) }
         { {\rm d}(g-r)\ {\rm d\,EW} }
    = \frac{1}{N_i} \sum_\nu \mathcal{N}_\nu(g-r, {\rm EW})
\end{equation}
where $N_i$ is the number of Voronoi regions in each galaxy, and $\mathcal{N}_\nu(g-r, {\rm EW})$ denotes the bivariate Gaussian corresponding to region $\nu$.
For a set of galaxies, the combined probability is simply $\Phi = \frac{1}{N} \sum_i \Phi_i$, where $N$ is the number of galaxies in the set, and the total fraction $f$ within a given domain is simply the integral of $\Phi$ over the corresponding area of the ageing diagram.
}

\subsubsection{Stellar mass and morphology}
Let us start by exploring the distribution of regions on the ageing diagram as a function of the total stellar mass and morphological classification.
For each bin in these global properties, we show in Figure~\ref{fig:distrib_morph_mass} the probability density distribution of all the regions summed over all galaxies within that bin.

The well-known relation between mass an morphology is clearly reflected on the preferential location of the galaxies across the diagonal of the main panels, that define a sequence that is mostly compatible with the ageing scenario, from primitive low-mass late spirals to evolved massive ellipticals.
It must be borne in mind, though, that such ageing sequence, analogous to the main stellar sequence in the Hertzsprung-Russell diagram, represents a snapshot at the present time rather than an evolutionary path: the progenitors of present-day ellipticals do not necessarily bear any resemblance to present-day spirals.
The latter are dominated by star forming processes, $f(\rm BE)\sim 1$.
As bulges become increasingly more prominent -- i.e. Sb's and early spirals --, the distribution of regions within the galaxies moves towards redder colours, $f(\rm RE) \gtrsim 0.25$.
At the end of the sequence, most regions in massive galaxies with lenticular or elliptical morphologies show red colours and mild or null \ha~emission, $f(\rm RA) \gtrsim 0.50$.
Very few regions, if any, are found in the BA domain throughout this ageing sequence.
In addition, it is worth noticing that late-type galaxies display a shallower probability distribution, extending across the whole BE region, while early-type galaxies are highly concentrated within the RA domain.

For late spirals, Sbc and Sb, total stellar mass drives the fraction of regions in BE to RE.
As galaxies become more massive, the contribution of the stellar bulge becomes more important; colours become redder, but there is always significant star formation activity within the galaxy.
Early spirals, in contrast, display a constant fraction $f(\rm RE)$, with decreasing number of BE regions in favour of the RA domain as mass becomes larger.
Elliptical and lenticular galaxies feature an inverse correlation between mass and BA fraction: from the (most common) massive to the lowest-mass systems, the BE fraction increases slightly, but the most significant change is that these galaxies become dominated by the BA domain.
In fact, these 45 galaxies with $M < 5 \times 10^{10}$~M$_\odot$ (out of 185 ellipticals and lenticulars), that amount to $\sim 7$ percent of the total sample, are the best candidates for hosting quenching processes.

In general, all panels show distributions in agreement with the results shown in \pp ~(grey solid lines), meaning that CALIFA measurements are enclosed within the same boundaries defined by the SDSS 90\% contour, with a few exceptions especially for low-mass ellipticals.
This is important, as it suggests that the physical processes driving the ageing of galaxies within the $3''$ aperture of SDSS observations extend to the regions covered by the FoV of CALIFA observations (up to $\sim 30$~kpc in our sample), and it is consistent with a common physical mechanism driving the `ageing' of galaxies both in the brightest central parts typically targeted by the SDSS fibres as well as in the much fainter outskirts considered in this follow-up work.

\subsubsection{Nuclear activity}

In addition, another process that has been proposed to play a major role in driving galaxy evolution is the presence of an Active Galactic Nuclei (AGN).
The vast amounts of energy that these physical processes generate, despite their small size and location, have been suggested as relevant feedback mechanisms and serious candidates for plausible quenching in galaxies \citep[e.g.][]{Sijacki+07,  Fabian12, Dubois+13, Zubovas+17, Sanchez+18, Lacerda+20, Kalinova+21}.

Figure~\ref{fig:distrib_bpt_ew} shows the distribution of Voronoi regions on the ageing diagram when galaxies are classified in terms of the BPT diagnostic diagram, supplemented by nebular emission equivalent width $\rm EW_{em}(H\alpha)$ as suggested by \citet[see Section~\ref{sec:global_properties} above]{Sanchez+21}.

As expected, panels corresponding to galaxies grouped as \emph{star-forming} display configurations mostly compatible with the ageing scenario.
As the nuclear $\rm EW_{em}(H\alpha)$ becomes smaller, the overall distribution shifts towards the RE domain.
On the other hand, \emph{composite} and \emph{AGN/Shocks} columns yield very similar results.
From high to low $\rm EW_{em}(H\alpha)$, the fraction of BE drops mainly in favour of the RA and BE domains, while RE remains roughly constant.
The distributions feature more irregular shapes than the \emph{star-forming} galaxies, departing sometimes from the contour obtained in\ \pp.
Nevertheless, we do not find strong evidences for quenching associated to nuclear activity.
The fraction of BA regions is always a minor contribution to the overall distribution, being largest when $\rm EW_{em}(H\alpha) < 3$~\AA.

\begin{figure*}
    \includegraphics[width=0.9\linewidth]{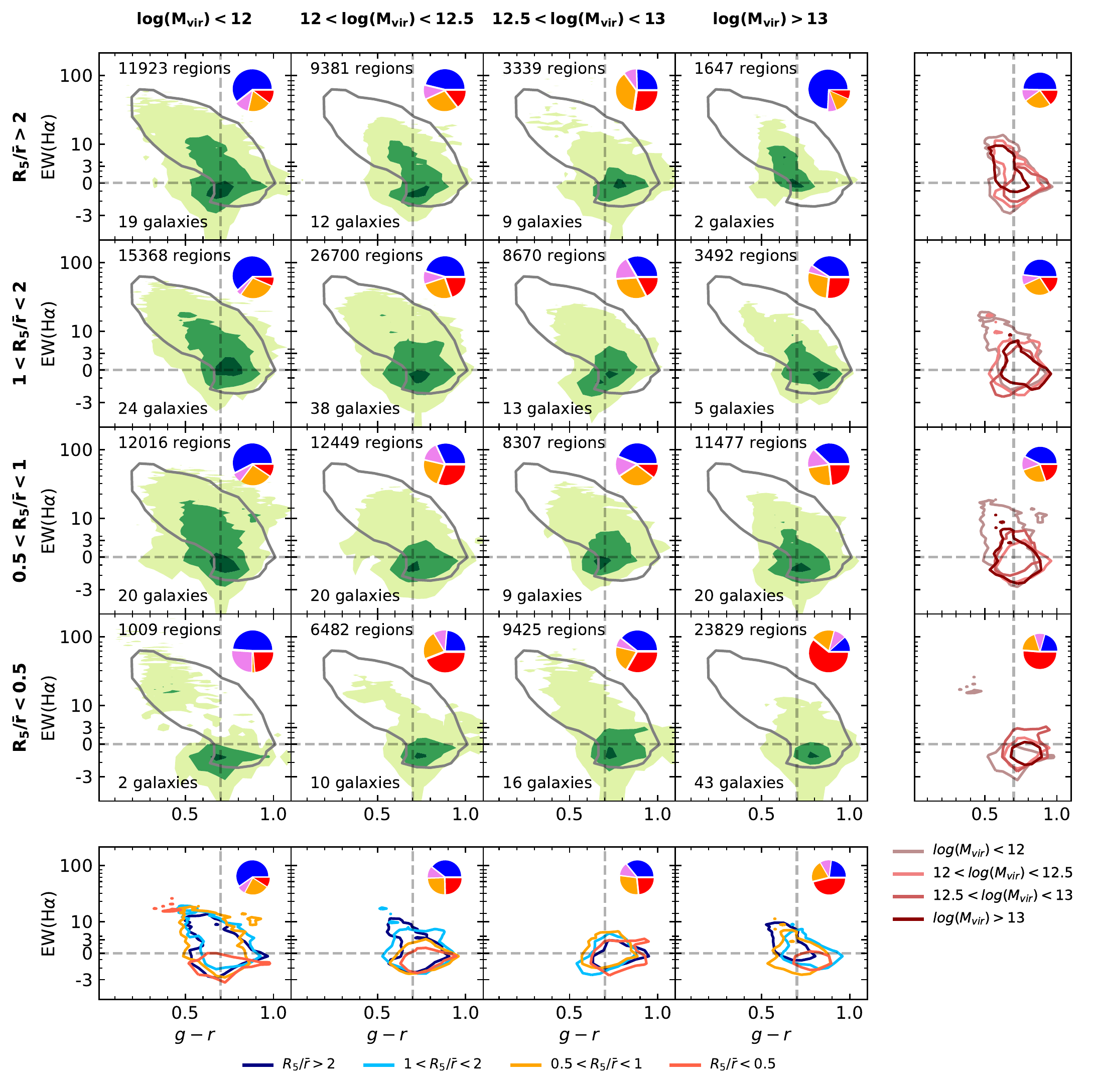}
    \caption{Dependence of the color-EW diagram on environmental properties.
    This figure is arranged similarly to Fig.~\ref{fig:distrib_morph_mass} but binned in terms of \envir~(rows), and group virial mass, $\rm M_{vir}$ (columns). }
    \label{fig:distrib_environment}
\end{figure*}
\subsubsection{Environment}

If environment were a driver of quenching \citep[e.g. ram-pressure stripping or strangulation;][]{Boselli+06, McCarthy+08, Ebeling+14, Peng+2015} it would be expected to leave an imprint on the ageing diagram.
Thus, we show in Fig.~\ref{fig:distrib_environment} the probability distribution of galaxy regions in different bins of group halo virial mass, $\rm M_{vir}$, and normalized distance to the fifth nearest neighbour, \envir. 

As illustrated before in Fig.\ref{fig:environment_sample}, the most populated bins correspond to the diagonal of the main panels, since both environmental proxies are well correlated.
Isolated objects in low-mass haloes are dominated by BE regions, albeit not as much as late-type spirals\footnote{We note that this morphological type is underrepresented in the \citet{Yang+07} catalogue.} of any mass.
Conversely, most of the regions in systems associated to dense environments and massive haloes are located in the RA domain.
The BA and RE fractions remain approximately constant, regardless of the galactic environment.

At fixed \envir, the probability distributions are typically shallower, with a larger fraction of regions in the BE domain, for low-mass haloes.
In contrast, the panels with $\log(M_{\rm vir})\gtrsim13$ display the largest $f\rm (RA)$ for any value of \envir.
The BE fraction is also sensitive to the galaxy overdensity as traced by this parameter.
For a given group halo mass, low-density regions tend to host galaxies with higher $f\rm (BE)$, while galaxies in dense environments present higher RA fractions.

Our results merely corroborate the well-known trends with galactic environment \citep[e.g.][]{Balogh+04, Baldry+06}, although they suggest that the correlation with the location on the ageing diagram is not as strong as for total stellar mass and galaxy morphology.
Furthermore, we do not find strong signatures for environmental quenching that are obviously associated to either of the environmental proxies under consideration.
The fraction of BA is roughly constant, $0.1\lesssim f(\rm BA) \lesssim 0.2$, across all panels.

\subsection{The role of local properties}
\label{sec:results_local_properties}
\begin{figure*}
    \includegraphics[width=0.9\linewidth]{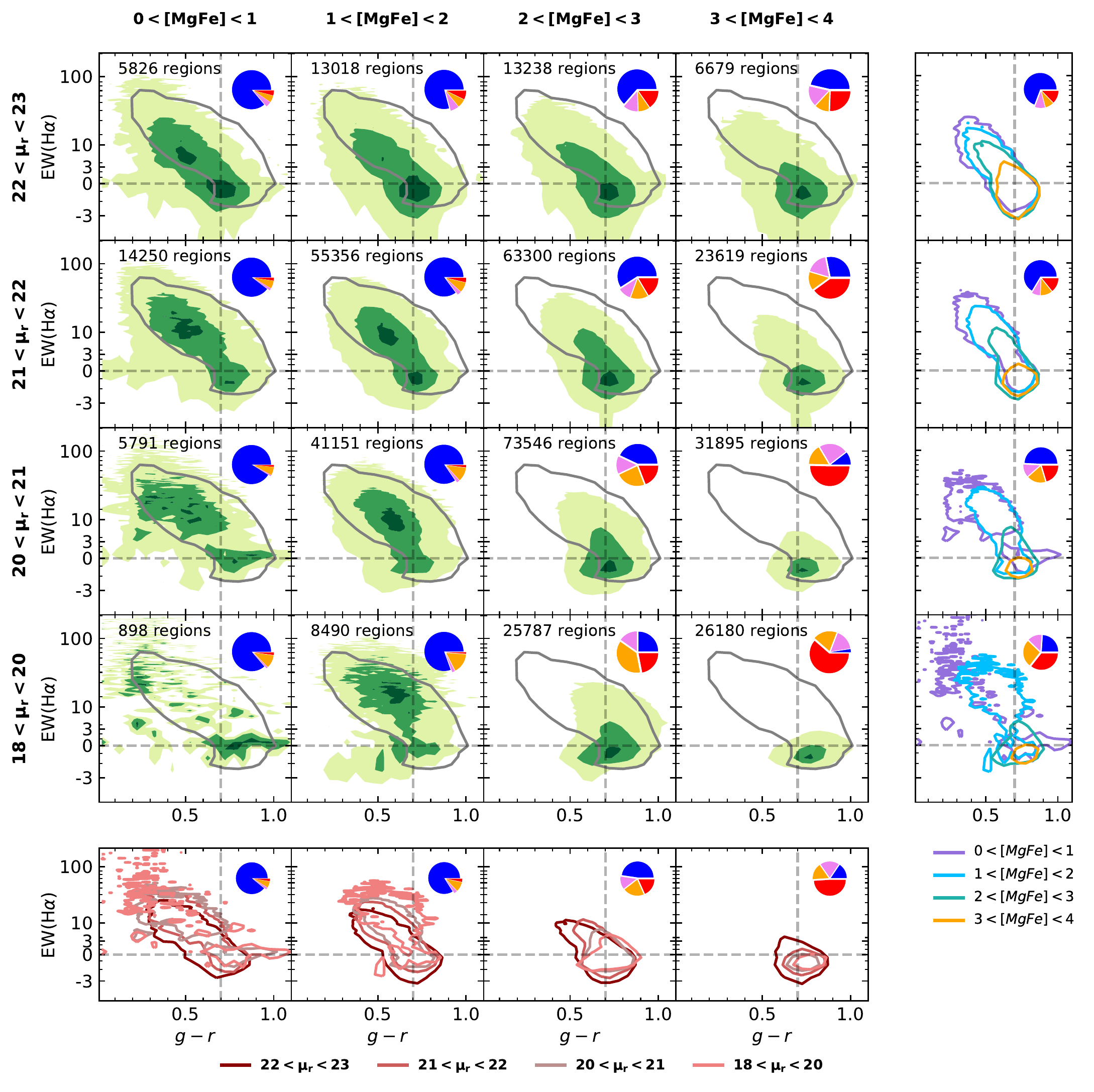}
    \caption{Dependence of the color-EW diagram on local properties.
    This figure is arranged similarly to Fig.~\ref{fig:distrib_morph_mass} but binned in terms of surface brightness (rows), $\mu_r$, and metallicity (columns), \MgFe. 
    \change{Each distribution is computed by summing all the Voronoi regions within a given bin.}}
    \label{fig:distrib_met_surface_brightness}
\end{figure*}

The effects of local properties on the recent star formation history are explored in terms of surface brightness $\mu_r$ (well correlated with stellar mass surface density and radial distance to the centre) and the metallicity-sensitive lick index \MgFe, a proxy for the region's chemical evolutionary stage and, indirectly, stellar-to-gas faction, as discussed in \citet[]{Ascasibar+15}.
Figure~\ref{fig:distrib_met_surface_brightness} shows the distribution of Voronoi regions on the ageing diagram in bins of $\mu_r$ and \MgFe, i.e. different regions within a galaxy will be located in different panels depending of their local properties, \change{and the probability distribution is estimated as $ \Phi = \frac{1}{N} \sum_\nu \mathcal{N}_\nu$, where the number $N$ of regions within each bin is indicated on the top left corner of every panel}.

In general, high values of \MgFe\ (i.e. more chemically evolved regions) are frequently found at the inner parts of the galaxy, while low values are more typical of the outskirts.
However, the correlation is not nearly as strong as the ones discussed above between mass and morphology or our two environmental proxies.
In this case, the regions are much more evenly distributed across the different panels, with only a few exceptions, very far away from the average relation between metallicity and surface brightness.
\change{Here we must caution the reader about the fact that \MgFe\ has also a non-negligible dependence on stellar age as well \citep[e.g.][]{Kuntschner+10}, implying that there may be a factor 2 or more difference in metallicity between the young and old part of the sequence for a given metallicity bin.} 
The fact that \MgFe\ is subject to significant uncertainties also contributes to the dispersion in this figure.

For a given column, the fraction of regions belonging to the BE domain of the ageing diagram tends to decrease with surface brightness, usually moving into the RE domain, and also towards the RA domain at high metallicities.
In fact, the effects of metallicity at any fixed $\mu_r$ are perhaps slightly more evident: $f(\rm BE)$ drops from low- to high-metallicity, mostly in favour of $f(\rm BA)$ and $f(\rm RA)$.

As can be seen at the right-most column, where all the distributions corresponding to a given surface brightness are put together, the contours of different stellar metallicity nicely trace different stages along the ageing sequence.
Although low-metallicity regions are scattered though the whole sequence with a shallow probability per unit colour and equivalent width, most of the area (and therefore the fraction of regions) belongs to the BE domain.
As metallicity increases, regions become significantly more concentrated towards the evolved end of the sequence at the RA domain.

We do not observe any obvious connection between quenching and local properties.
Albeit $f(\rm BA)$ tends to increase systematically with \MgFe, it never raises above $\sim 25$ per cent.

\section{Discussion}
\label{sec:discussion}

\begin{figure*}
    \centering
    \includegraphics[width=\linewidth]{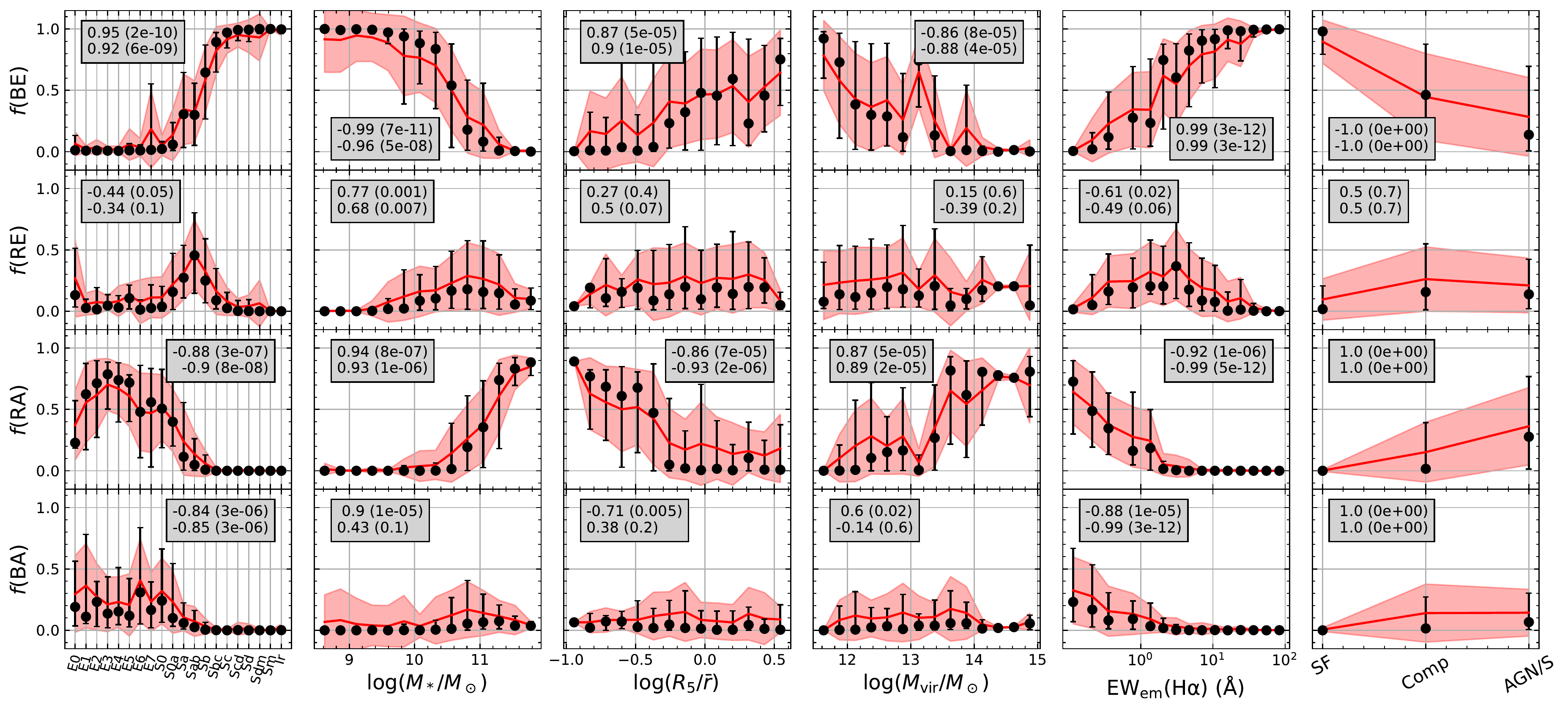}
    \caption{Domain's fraction as function of global properties. 
    Mean values with standard deviations and 50 percentiles with 84 and 16 percentiles are denoted by red lines and black points, respectively.  
    Grey boxes include the Spearman's rank correlation coefficient and p-value computed with the median (top) and mean (bottom) values, respectively.}
    \label{fig:correlations_global}
\end{figure*}

\begin{figure}
    \centering
    \includegraphics[width=1\linewidth]{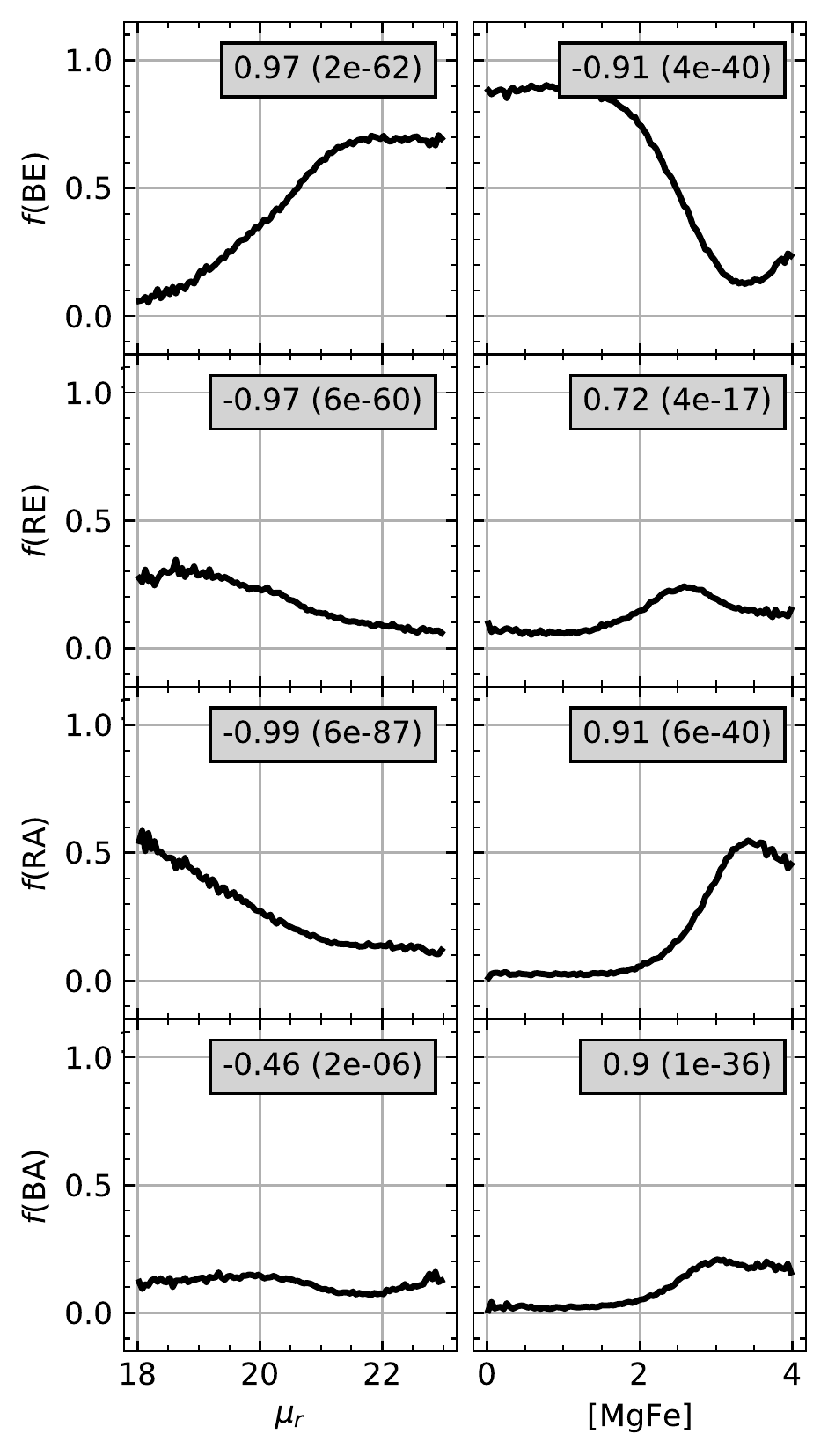}
    \caption{Domain's fraction as function of local properties. 
    Black lines represent the fraction of regions within each domain as function of surface brightness and [MgFe].
    Grey boxes include the Spearman's rank correlation coefficient and p-value.}
    \label{fig:correlations_local}
\end{figure}


In \pp, we interpreted the existence of a relatively narrow distribution in the colour-EW diagram (grey contours in Figures~\ref{fig:distrib_morph_mass}, \ref{fig:distrib_bpt_ew}, \ref{fig:distrib_environment}, and~\ref{fig:distrib_met_surface_brightness}) as a signature of a dominant, secular evolutionary mode in galaxies acting on timescales of the order of $\sim$Gyr.
We also proposed that any change in the star formation taking place on shorter timescales ($\lesssim 300$~Myr), i.e. a rejuvenation or quenching episode, would lead to a steeper trend in the diagram, as it would modify the EW(\ha) much faster than it would alter the colour (see Figure~\ref{fig_ageing_schema}).
We found little evidence of quenching in the SDSS sample, always associated to objects that display early and unknown morphologies located in dense environments. 

The results presented in the previous section confirm that objects classified as late spirals and Sbc's show trends fully compatible with long `ageing' timescales.
The rest of the spiral population (Sb's, and Early Spirals) would be compatible with faster evolution, but not necessarily quenching (which would occur on an even shorter timescale).
In addition, our results seem to indicate that nuclear activity and environment do not drive any sudden truncation of the recent star formation history of galaxies, at least on global scales.
Galaxies hosting AGN or living in denser environments are preferentially located at the RA end of the ageing sequence, hinting for accelerated evolution, but we do not observe an increased fraction in the BA domain.

We therefore advocate for \emph{ageing} being the natural driver of galaxy evolution for the majority of objects in our sample.
Here we will now explore the role of global and local properties in more detail, focusing not only on the statistical distribution of regions over the four domains domains defined in \S \ref{sec:results}, but also on the resolved ageing diagram of individual galaxies.
Finally, some caveats about systematic uncertainties will be discussed. 

\subsection{Global and local properties}

Fig.~\ref{fig:correlations_global} shows that all the global properties considered in the present work display a clear correlation with the domain fractions.
It is worth to note that the most-sensitive domains (i.e. those with the highest correlation coefficients for all proxies) are BE and RA.
In particular, the fraction of regions in these two domains seems to be bimodal in terms of morphology and total stellar mass, with values close to 100 per cent in the late-type or low-mass regime and vice versa.
The intermediate population, featuring the highest $f(\rm RE)$, mostly corresponds to spiral galaxies with prominent bulges ($\sim$Sab) in the mass range $10<\log(M_*/M_\odot)<11$, $\rm EW_{em}(H\alpha)\sim3$
and \emph{composite} nuclear classification.
The BA domain, expected to be populated by recently quenched regions, shows  weak trends with the majority of properties.
Recent quenching episodes seem to be more related to morphological classification and nuclear $\rm EW_{em}(H\alpha)$ values.
On the one hand, early-type galaxies show higher values of \ba, although presenting a large dispersion. 
Note that whilst environment shows mild correlations with \ba\, it is closely connected to galaxy morphology as seen in Figure~\ref{fig:environment_sample}.
Whether galaxy morphology is driving the process or a consequence of it is nevertheless an important (but yet unsolved) question.
On the other hand, galaxies with significant fractions of regions in the BA domain usually present nuclear $\rm EW_{em}(H\alpha)\sim 0$.
This symptom supports the inside-out evolutionary trend observed in the vast majority of galaxies in the CALIFA survey \citep[e.g.][]{Gonzalez-Delgado+17} and argues against a major role of AGN activity being responsible for quenching star formation on galactic scales \citep[][]{Ellison+21}, unless they are different evolutionary stages, as argued by e.g. \citet{Harrison+17}.
Finally, galaxies also show a mild increment of $f(\rm BA)$ when $\log(M_*/M_\odot)\sim 11$, roughly coincident with the transition from blue to red systems.

Our proposed environmental proxies present the lowest values of the correlation coefficients.
While it is well-known that environment plays a critical role regarding galaxy evolution \citep[e.g.][]{Boselli+06}, star formation is affected on timescales of the order of several gigayears \citep[e.g.][]{Peng+2015, Zinger2018}.
In fact, processes like strangulation, galaxy harassment or thermal evaporation, among others, might simply foster galaxy evolution on shorter timescales without sudden interruptions on the star formation history.
If gas accretion is interrupted, or is less efficient, one would expect that galaxies `age' more rapidly as they consume their remaining gas.
This can be translated into the ageing diagram as most galaxies transiting from BE to RA through RE rather than the BA domain.
Only systems undergoing strong ram pressure stripping in very dense environments ($M_{\rm vir}\sim10^{15} M_\odot$), with characteristic timescales of the order of $10^{7}-10^{8}$ yr \citep[e.g.][]{Kapferer+09, Ebeling+14}, could be identified as quenching events.
Our results are compatible with the idea of strangulation and/or rapid gas exhaustion in some galaxies.
Nevertheless, we must caution the reader that galaxies living in clusters are scarcely represented in our sample, preventing us from making any robust claim in this regard.
Other galaxy surveys such as SAMI \citep[][]{Bryant+15}, specifically targeting different environments, are better suited to shed light on the impact of nurture on galactic ageing and quenching on global or resolved scales.

Surface brightness and stellar metallicity are found to correlate with most domain fractions as shown in Figure\ \ref{fig:correlations_local}.
High values of $\mu_r$, typically corresponding to inner regions, tend to present low values of \be\ in favour of RA and RE. 
This is again consistent with \change{previous works \citep[e.g.][]{Gonzalez-Delgado+15, Zibetti+17} that explored the correlation between surface brightness and mean stellar age.
In particular, our results based on the ageing diagram corroborate that brighter regions, often in early-type galaxies, tend to host older stellar populations, while the opposite is true for low surface brightness regions, typical of spiral-like systems.}

\change{
However, we do not find evidence for bimodality in the ageing diagram, as reported by \citet{Zibetti+17} for the luminosity-weighted age with respect to stellar surface density on resolved scales.
In the ageing diagram, all stellar populations older than $\sim 1$~Gyr cluster around $(g-r) \sim 0.75$ and $EW(\ha) \sim -1$~\AA, whereas younger regions scatter mostly through the BE domain.
Therefore, we find that the overall statistical distribution of all regions peaks in the RA domain, close to the centre of our classification lines, with a tail towards BE whose prominence depends on the adopted weighting scheme.
As noted in \citet{CorchoCaballero+20}, the shape of the probability distribution, including the number of peaks, is sensitive to the adopted variables and their systematic and statistical uncertainties.
Even if two peaks are found, that does not imply that the physical properties of galaxies or regions are intrinsically bimodal in any sense.
Rather than the traditional view, where there is a universal bimodal distribution of star formation histories \citep[`main sequence' and `quenched'; see e.g.][]{Corcho-Caballero+21}, whose peaks are unevenly populated by different galaxies, we would like to support the alternative interpretation in terms of a gradual ageing sequence that maps non-linearly to different sets of variables.
We consider that this is an open issue, and further work is required in order to discriminate between both scenarios.
}

\change{Our results are consistent with} an inside-out ageing scenario for the majority of galaxies.
Approximately 70 per cent of low surface brightness regions are found in the BE domain. 
The fraction of potentially quenched regions, \ba, does not depend remarkably on surface brightness.
Fractions are always equal to or below 10-15 per cent for a given $\mu_r$.
In terms of stellar metallicity, 90 per cent of the regions with $\rm [MgFe]^\prime\leq 2$ belong to BE, with the remaining 10 percent roughly located at RE.
It is only at high metallicity ($\rm [MgFe]^\prime\geq 3$) that \ba\ accounts for 20 per cent of the regions.

The physical process(es) that we are observing seem thus to depend on local properties, while simultaneously there is an evident correlation with galaxy mass and morphological type, as well as, to a lesser extent, environment and nuclear activity.
As a matter of fact, our three global properties are known to correlate with each other, and it is thus complicated to determine which one of them is driving the other.
In any case, our results seem to support a scenario where global properties determine the evolution of the galaxy as a whole, whereas it is the local properties that determine how such evolution takes place within the system.

\subsection{Ageing and quenching within individual galaxies}


We argued before that local processes, traced by the surface brightness, may play an important role in controlling the process of `ageing' in galaxies.
The distributions of Voronoi regions in the colour-EW diagram are plotted on Figure~\ref{fig:all_figs} for each galaxy in the sample. 
They are elongated and present radial trends consistent with the inside-out galaxy formation scenario \citep[e.g.][]{Tacchella+15, Gonzalez-Delgado+16, Belfiore+17, Sanchez+18, Woo&Ellison19, Bluck+19, Sanchez+20}, as illustrated by the maps of the spatial location of the BE, RE, BA and RA domains.

At variance with the SDSS data used in \pp, IFS observations reveal that the location in the colour-equivalent width diagram of all the regions within a single object is typically not concentrated, but broadly extended along the ``ageing sequence''. 
Outer regions tend to display in general larger values of EW(\ha) and bluer colors than the inner counterparts.
This result strengths the hypothesis of `ageing' being a \emph{very local process}, as well as quenching or, in a more general sense, star formation processes.  
Consequently, it emphasizes the relevance of considering aperture effects when dealing with single fibre data.

However, the diversity of distributions in the individual diagrams suggests that there may be different `ageing' tracks.
Objects with `earlier morphologies' (in particular, lenticulars and ellipticals) seem to have finished such evolutionary phase, as the vast majority of their Voronoi regions are located in the evolved RA end of the `ageing sequence''.
CALIFA data show that the transition between the centre and the outer parts of galaxies is continuous in the colour-equivalent width diagram.
Such radial trend is present for every galaxy in our sample (although it is most obvious among Sb, Sab, and Sa galaxies), and it supports the idea that `ageing' is a universal mechanism affecting all galaxies of all types over their entire extent.
In particular, this mechanism is not restricted to galaxies undergoing a transitional process from a `star-forming' to a `passive' state, but it is the normal mode of evolution throughout cosmic history.

We also find some rarer systems for which inner regions present larger values of EW(\ha) than the outer parts (e.g. NGC0693).
This trend, predominantly found for spiral systems, may be produced by the presence of central bars \citep[e.g.][]{Lin+17} or galaxy interactions \citep[e.g.][]{Li+08} that funnel cold gas to the nuclear regions enhancing thus the star formation rates.

Although `ageing' seems to proceed faster in high stellar density regions, which arguably favour the accretion and consumption of gas, we do not find a universal relation between evolutionary state and local surface brightness.
Quite the contrary, our results demonstrate that the resolved colour-equivalent width diagrams show a wide variety of distributions, even at fixed surface brightness. 
While the regions of late spirals are narrowly located along the `chemically-primitive' part of the sequence, objects classified as lenticulars (S0a and S0) and ellipticals predominantly gather near the `evolved' end, showing most of their \ha\ in absorption.
The rest of the objects (Sb's, Sab's and Sa's) show a `mixed' distribution in the diagram (elongated across the colour-EW plane), compatible with some parts of them actively forming stars, some other with negligible star formation, and the vast majority showing mild activity (with $0<$EW(\ha)$<10$~\AA) and intermediate colours.
If the secular conversion of gas into stars (ageing) is the main driving mechanism of the evolution of a galaxy, it should not be surprising to find these continuous distributions in the colour-EW diagram, i.e. the existence of this population of `mixed' galaxies (which, in fact, dominate our sample, and are not unlike our very own Milky Way) must be fairly common in the Universe.

Nevertheless, we also find distributions of galaxy regions compatible with having experienced recent quenching episodes. 
Out of the 637 objects in our sample, 84 ($\sim13\%$), 40 ($\sim6\%$) and 19 ($\sim2\%$) of them present $f(\rm BA)$ larger than 25, 50 and 75 per cent, respectively. 
As mentioned before, quenching seems to be correlated with mass and morphology, since early type systems predominate over late type galaxies.
For example, regarding those galaxies with $f(\rm BA)\ge 0.25$ only 13 present Sa (5), Sab (5), Sb (2) and Sbc (1) morphologies.
On the other hand, we find that low mass elliptical galaxies have a much higher probability of populating the BA domain at the present time than massive ellipticals.

\subsection{Systematic uncertainties}

The ageing diagram provides a model-independent tool to identify sudden changes in the recent star formation history of galaxies.
However, it is important to note that the proxies used to trace the presence of young and intermediate age stellar populations might also be sensitive to other factors.

As mentioned above, young stars are not the only ionizing source in galaxies. 
Hot evolved stars such as HOLMES or post-AGB \citep{BInette+94, Flores-Fajardo+11} may also produce hard radiation fields, as well as shocks coming from galactic winds \citep{Heckman+90, Veilleux+05} or stellar remnants related to previous SF processes \citep{Sanchez+12b}. 
In addition, another possible `local' process that has been proposed to play a major role in driving galaxy evolution is the presence of an Active Galactic Nuclei (AGN).
The vast amounts of energy that these physical processes generate, despite their small size and location, have been suggested as relevant feedback mechanisms and serious candidates for plausible quenching in galaxies \citep[e.g.][]{Sijacki+07,  Fabian12, Dubois+13, Zubovas+17}.
The presence of ionising sources other than O and B stars might produce an overestimation of the EW(\ha), potentially shifting the affected galaxies or regions vertically in the ageing diagram, from the absorption towards the emission domains.


On the other hand, optical colours such as $g-r$ may be substantially affected by dust extinction. 
Typically, color excesses on resolved regions of CALIFA galaxies range between $E(g-r)\sim 0.1-0.3$ magnitudes, which is clearly sufficient to move an intrinsically blue region to the red domains of the diagram.
This caveat may be overcome by explicitly modelling (differential) dust extinction and recovering the recent star formation history by means of evolutionary population synthesis models \citep[e.g][]{Heavens+00, Cid-Fernandes+05, Ocvirk+06, daCunha+08, Sanchez+16, Leja+17}.
While this is indeed a powerful technique, that would in principle make possible to directly estimate the ratio between the star formation rate in two arbitrarily chosen timescales, it is based on a much larger set of assumptions than the model-independent approach provided by the ageing diagram.
Having different systematic biases, we argue that both of them should be regarded as complementary.

The location of a given region on the ageing diagram is also sensitive to the metallicity of both the stellar population and the diffuse interstellar gas, affecting colours as well as the intensity of emission and absorption lines.
This might be, in fact, related to the diversity of trends we observe in different individual galaxies.
Such diversity, in any case, hints that more parameters are indeed needed in order to fully characterise the spectral energy distribution.
Metallicity, dust abundance, gas fraction, and recent star formation are nevertheless well known to be strongly correlated \citep[e.g.][]{Mannucci+2010, Lara-Lopez+10FP, Lara-LopezLSH13, Lilly+13, Ascasibar+15, Barrera-Ballesteros+20}.
Theoretical models will be used in future work to identify the statistically optimal indicators (e.g. Lick indices, multi-wavelength photometry, kinematics) to extend the ageing diagram and break the degeneracy between all these factors.

\change{As noted before, galaxy distributions in the ageing diagram were directly computed by summing all Voronoi regions brighter than $\mu_r=24~\text{mag/arcsec}^2$. 
We find that the results obtained using bare Voronoi regions are similar to the ones obtained when weighting each Voronoi bin by its total luminosity in the $r$ band, proxy for the total stellar mass within that region.
This arbitrary choice yields distributions typical of the bulk of stellar mass formed in galaxies, but it might not be representative of the galaxy outskirts since the number of low surface brightness spaxels is underrepresented. 
In order to effectively include the outer rims of galaxies in the ageing diagram, it could be useful to consider alternative schemes such as weighting every Voronoi region in terms of their corresponding physical projected area, at the price of introducing more noise in the results. 
Nevertheless, this kind of analysis lies beyond the scope of the present work as we aim to provide the current evolutionary state (as illustrated by the domains fractions) for each galaxy in the sample within the context of studying variations in terms global and local properties.
}
\subsection{\change{Sample bias}}

\change{
CALIFA is not a volume nor flux limited survey, but a diameter-selected sample.
The selection function is thus not trivial, but it has been carefully modelled by \citet[]{Walcher+14} in order to derive appropriate volume corrections for a rigorous statistical study.
Unfortunately, there is a significant population of galaxies (most notably, at low-mass end) that are not adequately represented.
The CALIFA extensions solve this problem by specifically targeting those objects, at the price of sacrificing the statistical characterisation of the selection function.
There are, in principle, two aspects that may be relevant for the results reported in the present work:
}

\change{
On the one hand, the $g-r$ colour distribution of our sample (Fig.~\ref{sample_global_SDSS}) does not exactly reproduce the results obtained by larger surveys of the nearby universe \citep[e.g.][]{Blanton+03}, although most differences are strongly mitigated (for the CALIFA mother sample, in the $M_r<-18.6$ range) after applying the \citet{Walcher+14} correction.
A more uniformly selected sample would of course be extremely useful in order to derive statistical distributions over the whole luminosity and colour range.
On the other hand, our results highlight that galaxy outskirts may occupy a very different region of the ageing diagram than the central parts, which makes CALIFA data ideally suited to explore this issue.
}

\change{
In terms of environment, Figure~\ref{fig:environment_sample} evidences a certain lack of (low-mass) satellite galaxies, even after including the extension sample, where these systems seem to be less represented than their isolated counterparts.
At high $M_{\rm vir}$, low-mass satellite galaxies should dominate the number counts, but the present sample mainly includes massive central galaxies.
This is also testified by the observed correlation between stellar and halo mass, which is known to be much tighter for central galaxies only \citep[see e.g.][]{Rodriguez-Puebla+12, Tinker+13}.
Consequently, environmental effects in this work might be underrepresented, and their net effect on the ageing diagram could become more evident when considering a more environmentally-complete sample.
}
\section{Summary and Conclusions}
\label{sec:conclusions}

We have analysed 637 galaxies from the CALIFA survey following the same methodology presented in \citet{Casado+15}. 
We studied the dependence of the distribution of spatially-resolved galaxy regions in the colour-EW(\ha) diagrams with global -- morphology, mass, environment and nuclear activity -- and local properties -- surface brightness and stellar metallicity.
The characterization of the SDSS sample conducted in \pp\ relative to the morphology, mass, and environment is consistent with other estimators in the literature used within the CALIFA collaboration \citep[measurements of group properties, derivation of stellar masses, and visual morphological classifications obtained from ][]{Yang+07, Walcher+08, Walcher+14}.

In order to quantify the statistical distribution of galaxy regions across the `ageing diagram', we compute the fraction of regions within the four quadrants defined by $(g-r)=0.7$ and EW(\ha)$=0$. \emph{Ageing} can be understood as a slow evolution from the blue emission (BE) to the red absorption (RA) domains, transiting over the red emission (RE) domain over several Gyr, whereas sudden \emph{quenching} of the star formation activity would imply a rapid transition through the blue absorption (BA) domain on a timescale of the order of $\sim 300$~Myr.

Our main results can be summarised as follows:

\begin{enumerate}

\item Galaxies `age' on resolved scales and over their entire extent, i.e. the evolution of most of their regions is consistent with secular processes without sudden interruptions of the star formation activity.

\item
Nevertheless, we do find a small number of galaxies (predominantly low-mass early-type systems) with a sizeable fraction of their extent in the BA domain, compatible with having experienced recent quenching episodes.

\item The distribution of the entire population (spaxels) of each object in the colour-EW diagram depends mainly on its mass and morphology, as well as (secondarily) on environment and nuclear activity. 
Spaxels/regions of isolated late-type objects populate the star-forming part of the sequence, while early-types dominate the red end. The `intermediate' objects (Sb's, Sab's and Sa's) show in some cases a steeper trend compatible with large differences in evolutionary stage between the central and the outer parts.
Nuclear activity seems to be weakly connected with the overall distribution of regions.
 There is no clear evidence that AGN has a strong impact on star formation neither on galaxies as a whole nor on the observed radial trends.

\item
Local surface brightness and stellar metallicity are also correlated with the location on the ageing diagram. In general, dense and chemically-rich regions tend to be located at the evolved RA domain, whereas the low-metallicity galaxy outskirts populate the primitive BE extreme.

\item Although surely affected by scatter, most of the regions considered for all objects fall within the 90\% contour obtained for the SDSS sample. 
However, we find evidence that there may be different paths through the colour-equivalent width diagram rather than a universal, intrinsically thin `resolved ageing sequence', merely broadened by measurement uncertainties.
 
\item Most trends observed in the resolved `ageing' diagram present a radial dependence (as interpreted from the surface brightness dependence). 
This result suggests that, whichever physical mechanism is responsible for driving the process of ageing, it behaves in an inside-out fashion. 
 
\end{enumerate}

In a nutshell, this follow-up work reinforces the `ageing' scenario proposed in \pp, suggesting that the universal (unavoidable) process of gradual conversion of gas into stars affects galaxies of all morphological types throughout their entire extent, from the bright bulges and star forming regions typically targeted with the SDSS fibres to the faint outskirts of galaxies.
However, the simple, model-independent approach provided by the domain classification of the ageing diagram may be missing a number of quenched regions that are significantly reddened and/or hosting ionising sources not associated with young stellar populations.
Further extensions of the ageing diagrams will be explored in future work in an attempt to overcome these biases.

\section*{Acknowledgements}

We thank the anonymous referee for a careful and constructive revision of the manuscript. In particular, the use of the full CALIFA sample as well as different statistical indicators has notably increased the robustness of our conclusions.

This work has been supported by the Spanish Ministry of Economy and Competitiveness (MINECO) through the MINECO-FEDER grants AYA2016-79724-C4-1-P, AYA2016-77846-P, and PID2019-107408GB-C42.
RGB acknowledges additional financial support from the State Agency for Research of the Spanish MCIU through the Center of Excellence Severo Ochoa award to the Instituto de Astrof\'isica de Andaluc\'ia (SEV-2017-0709), grants PID2019-109067GB-I00 (MCIU), and P18-FRJ-2595 (Junta de Andaluc\'ia).

This study uses data provided by the Calar Alto Legacy Integral Field Area (CALIFA) survey (http://califa.caha.es/). Based on observations collected at the Centro Astronómico Hispano Alemán (CAHA) at Calar Alto, operated jointly by the Max-Planck-Institut fűr Astronomie and the Instituto de Astrofísica de Andalucía (CSIC).

This research has made use of NASA's Astrophysics Data System Bibliographic Services.

\section*{Data availability}
The data underlying this article are available at \url{https://classic.sdss.org/dr7/} for the SDSS survey and \url{https://califa.caha.es/} for the CALIFA survey. Additional data generated by the analyses in this work are available upon request to the corresponding author.
 \bibliographystyle{mnras}
 \bibliography{references}


\appendix
\section{Individual distributions.}
\label{appendix:individual_distributions}

The following figures provide the distribution of Voronoi regions in the color-equivalent width diagram for each galaxy in the sample, coloured in term of their surface brightness. 
In addition, we have also computed RGB composite images based on $gri$ bands following \citet[]{Lupton+04} and maps of domain spatial distribution.

\begin{figure*}
    \centering
    \includegraphics[width=\linewidth]{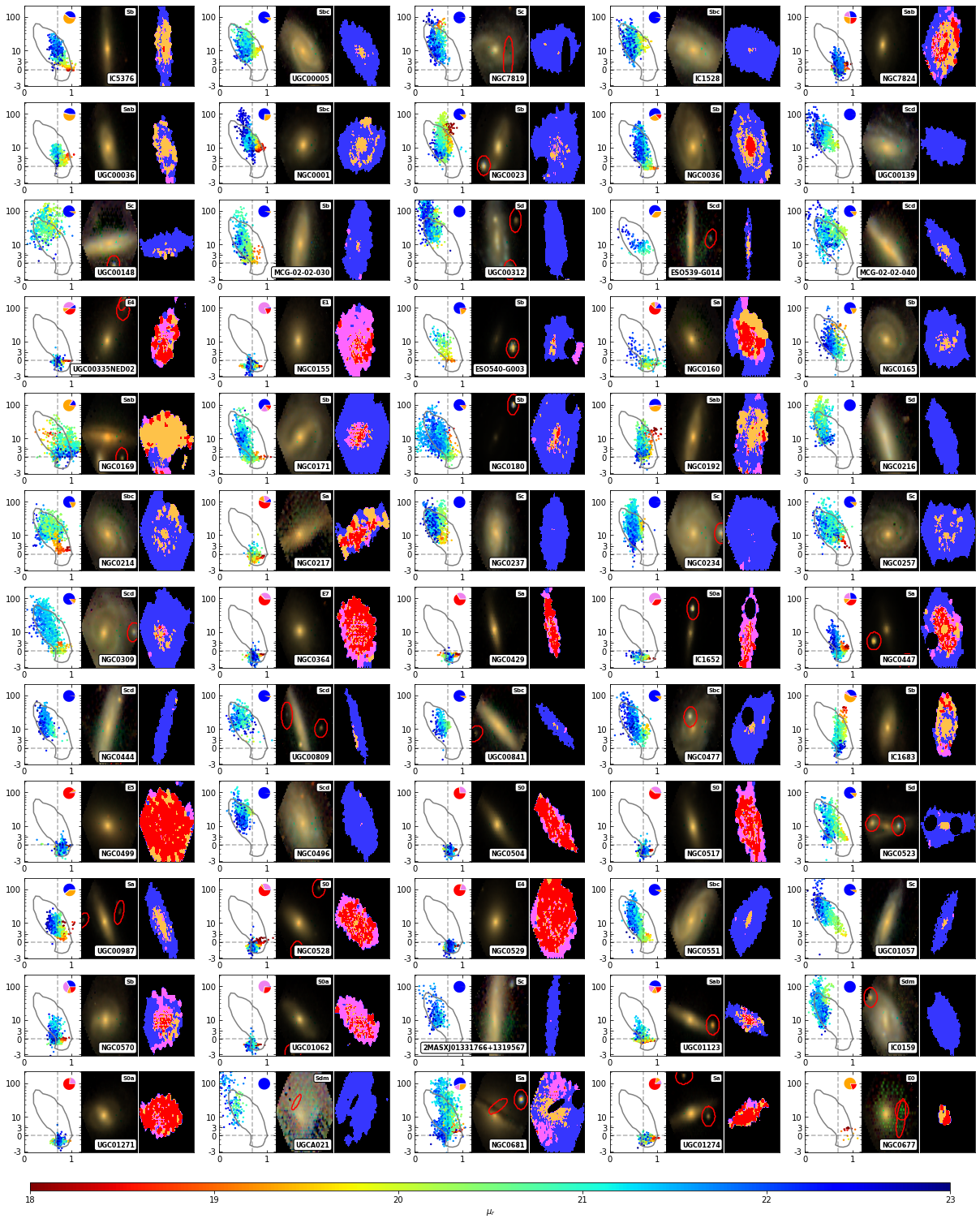}
    \caption{Individual galaxy distributions. 
    The distribution of Voronoi regions on the ageing diagram is represented on left panels for each galaxy with colours denoting surface brightness. 
    A pie chart indicates the fraction of regions within each domain: blue-emission (blue), red-emission (orange), blue-absorption (pink) and red-absorption(red).
    Central and right panels denote RGB composite images (including catalog name and morphological classification)-- based on the $g$, $r$ and $i$ bands -- and spatial distribution of domains, respectively.} 
    
    \label{fig:all_figs}
\end{figure*}

\begin{figure*}
    \centering
    \includegraphics[width=\linewidth]{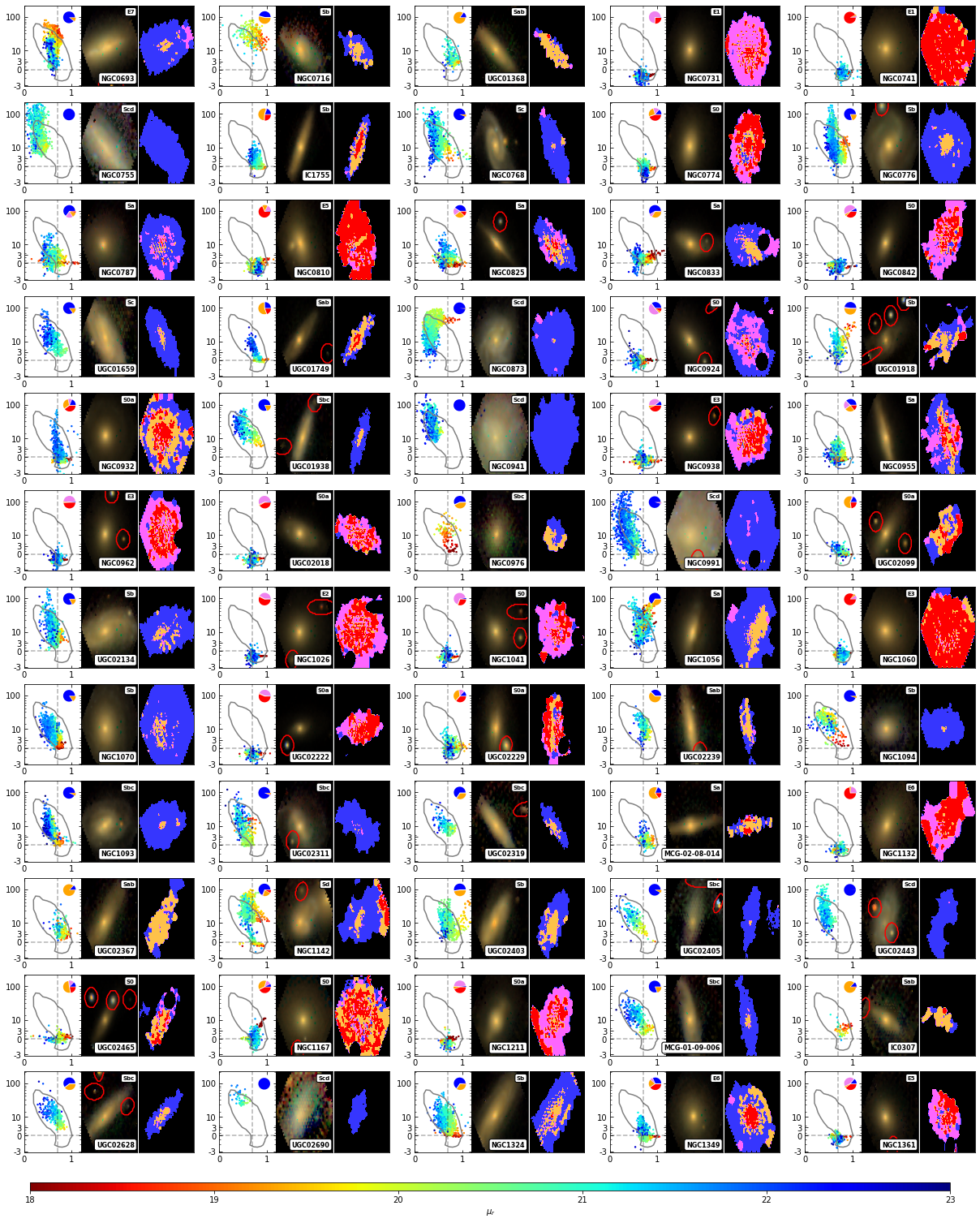}
    \contcaption{}
\end{figure*}

\begin{figure*}
    \centering
    \includegraphics[width=\linewidth]{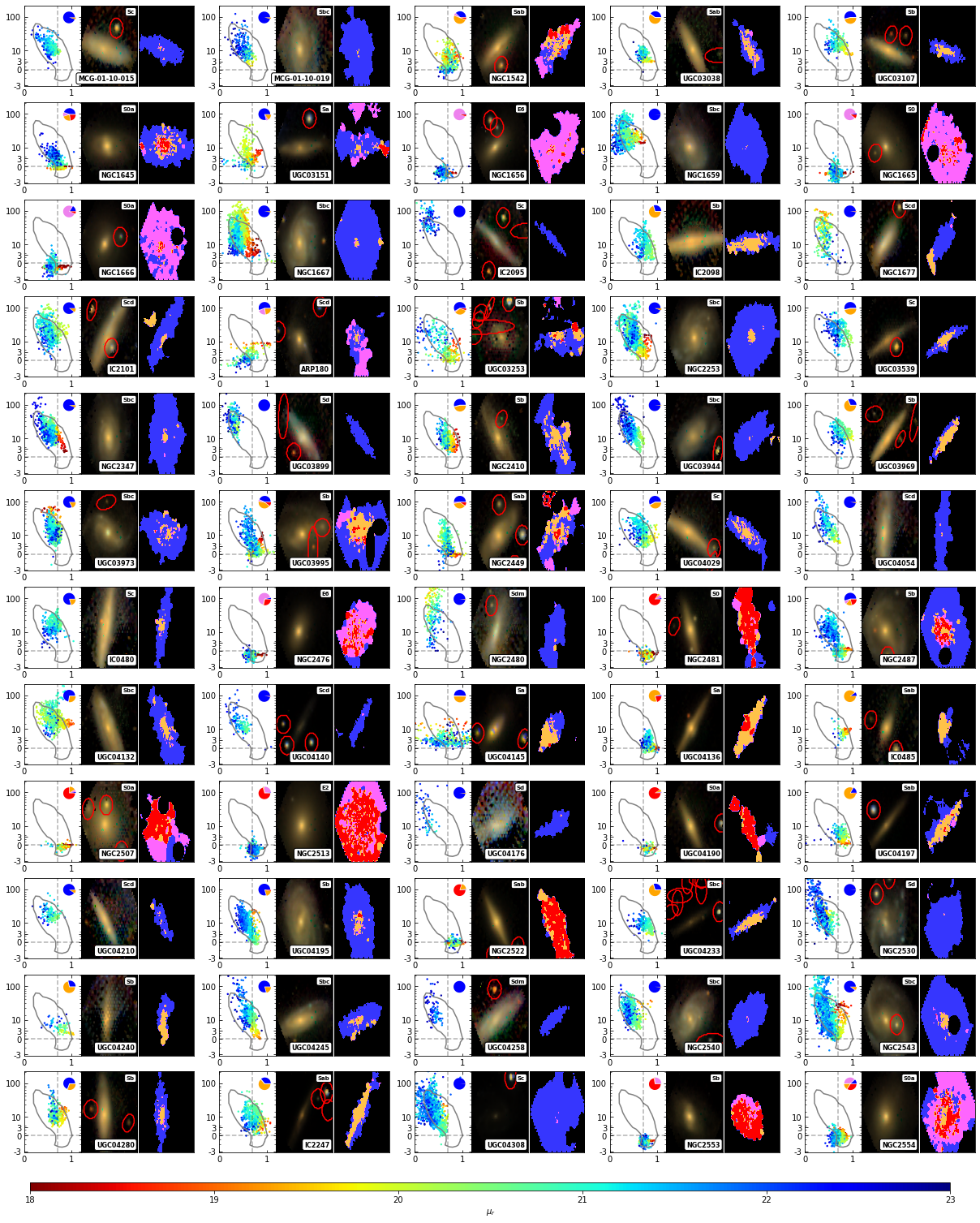}
    \contcaption{}
\end{figure*}

\begin{figure*}
    \centering
    \includegraphics[width=\linewidth]{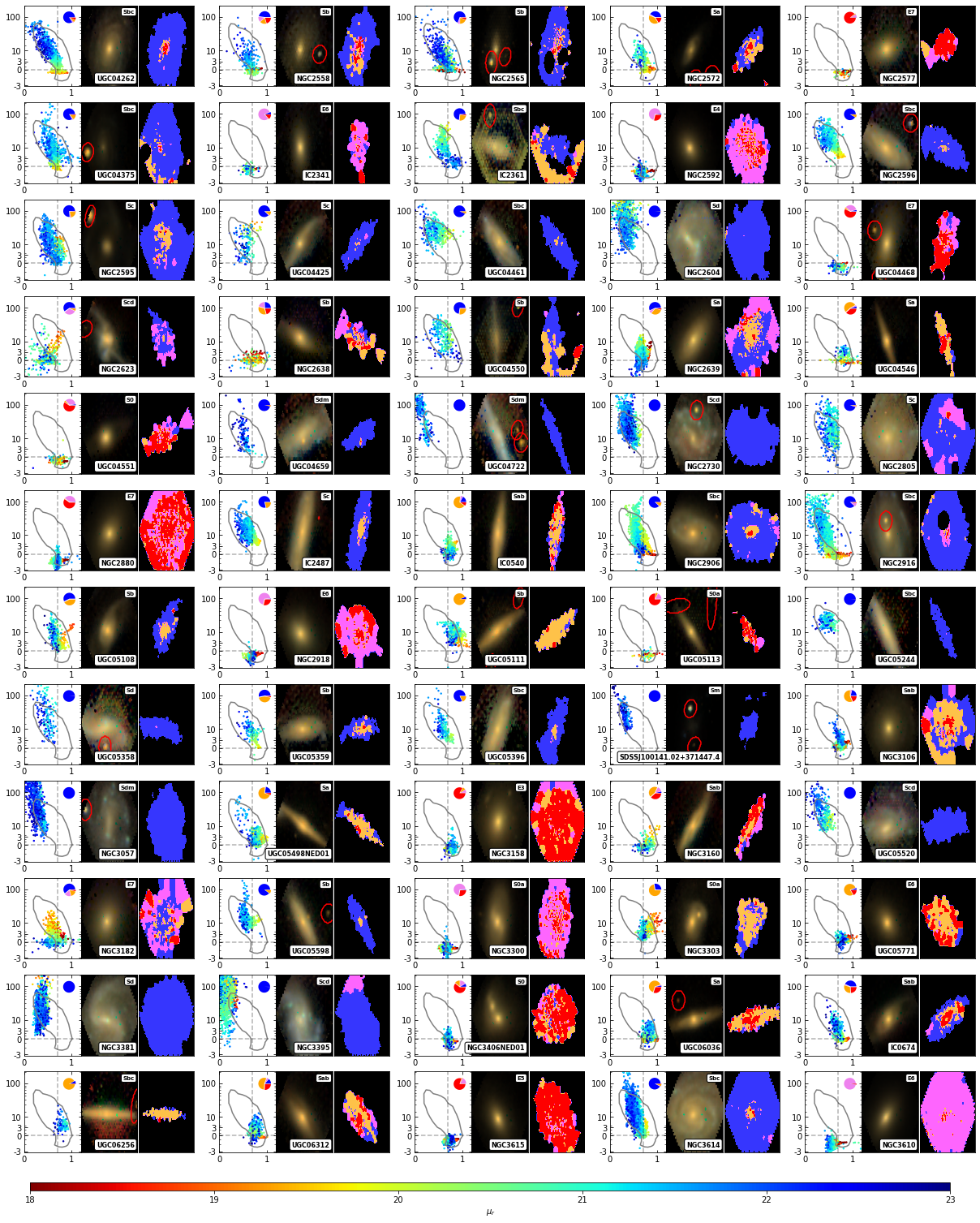}
    \contcaption{}
\end{figure*}

\begin{figure*}
    \centering
    \includegraphics[width=\linewidth]{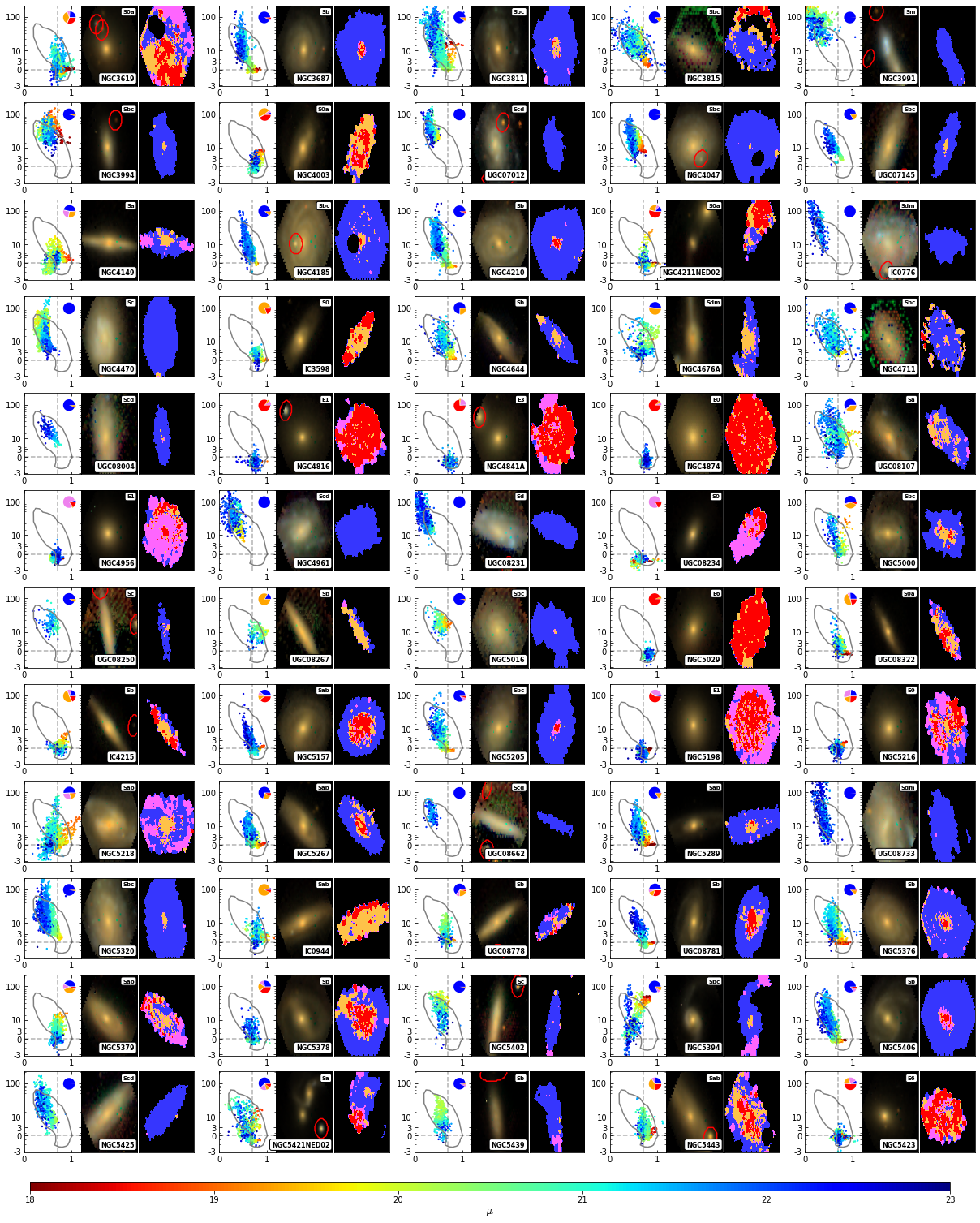}
    \contcaption{}
\end{figure*}
    
\begin{figure*}
    \centering
    \includegraphics[width=\linewidth]{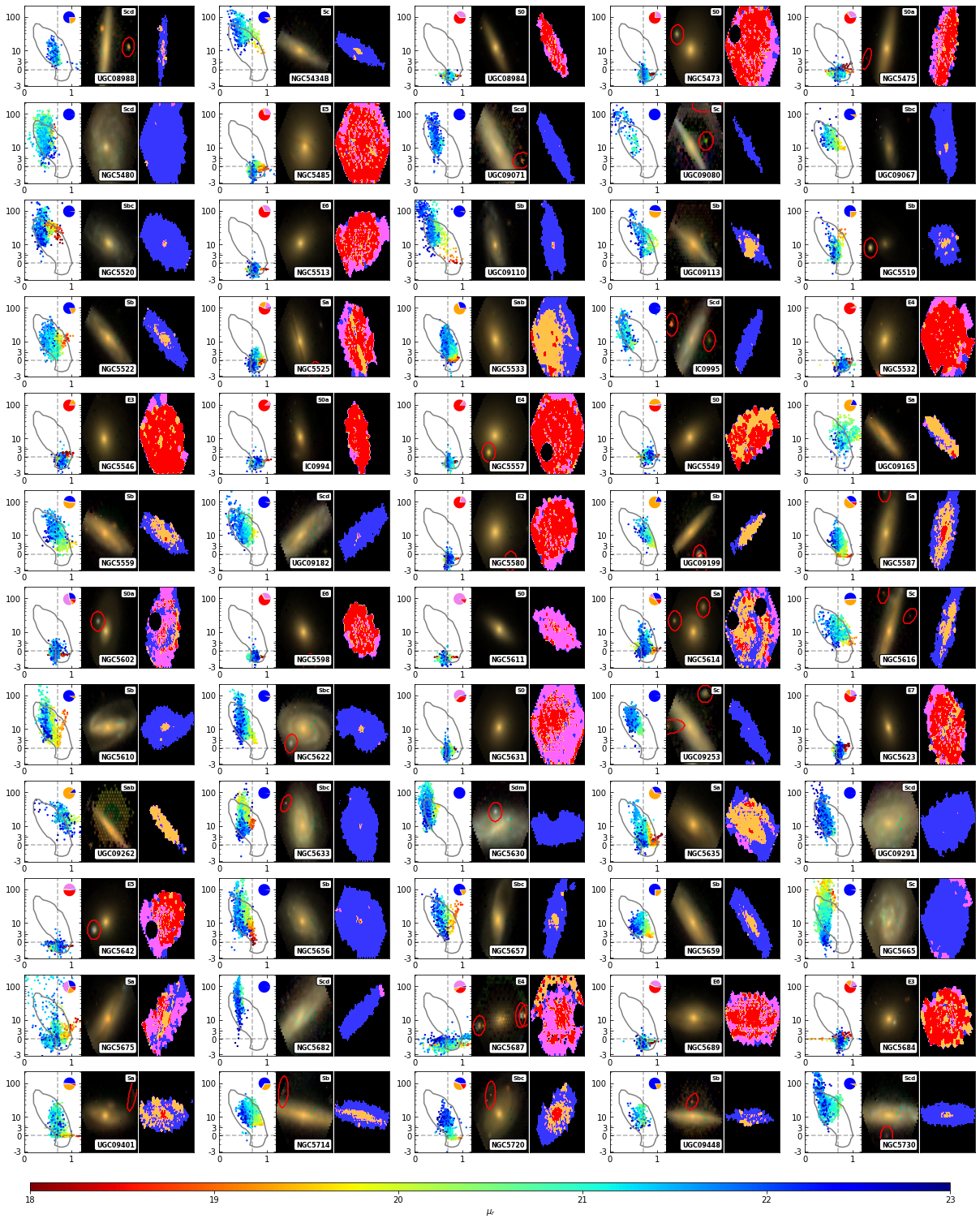}
    \contcaption{}
\end{figure*}

\begin{figure*}
    \centering
    \includegraphics[width=\linewidth]{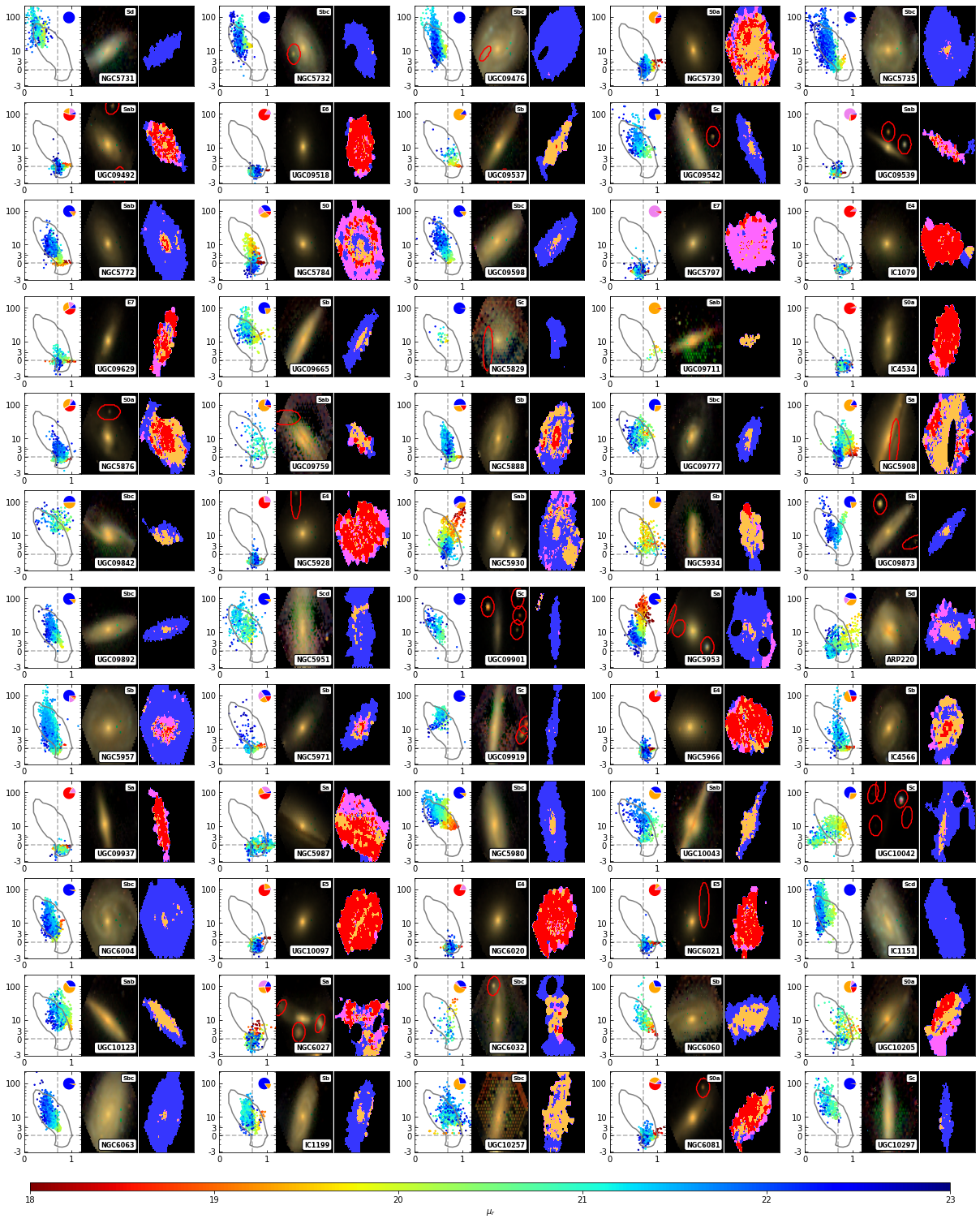}
    \contcaption{}
\end{figure*}

\begin{figure*}
    \centering
    \includegraphics[width=\linewidth]{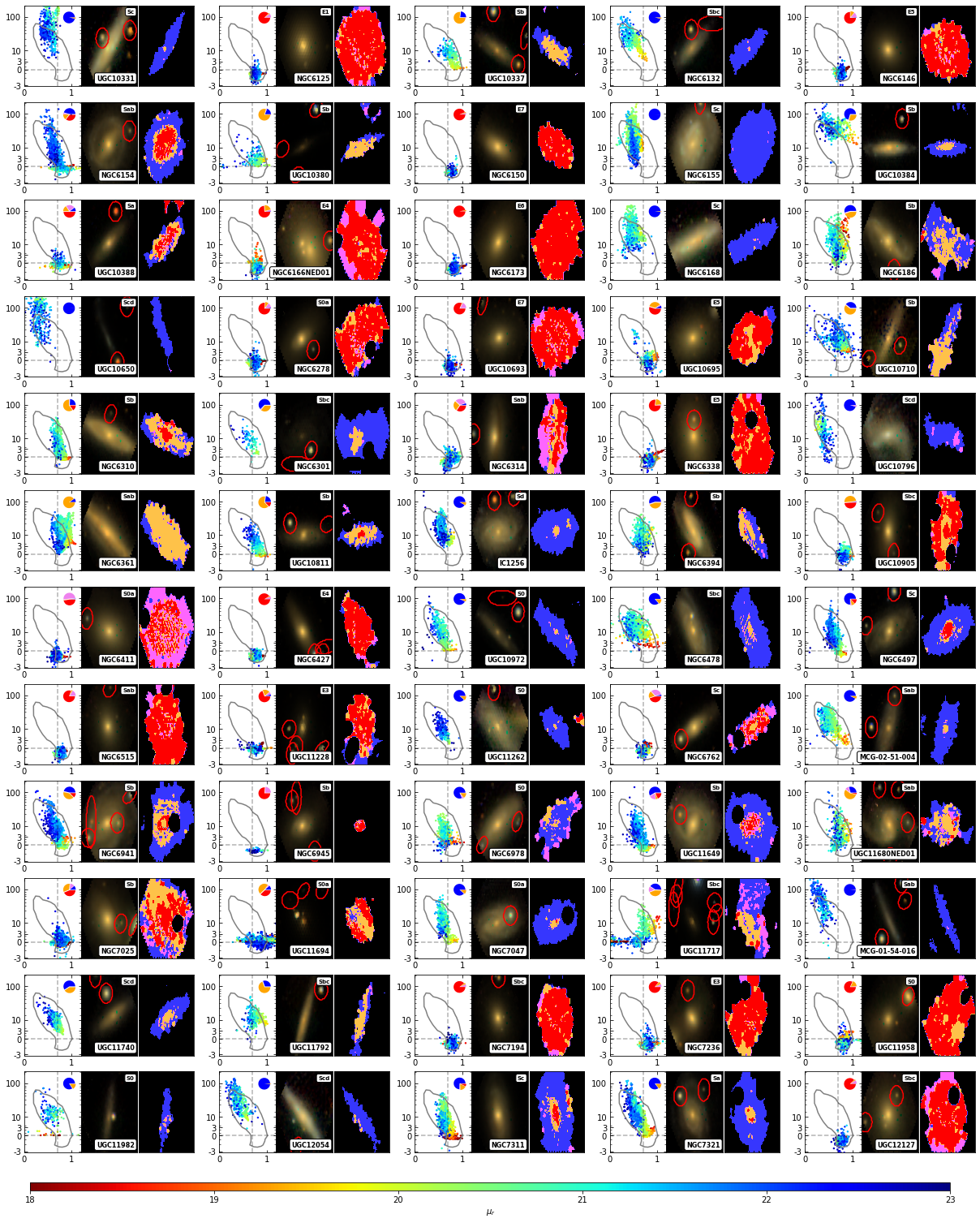}
    \contcaption{}
\end{figure*}

\begin{figure*}
    \centering
    \includegraphics[width=\linewidth]{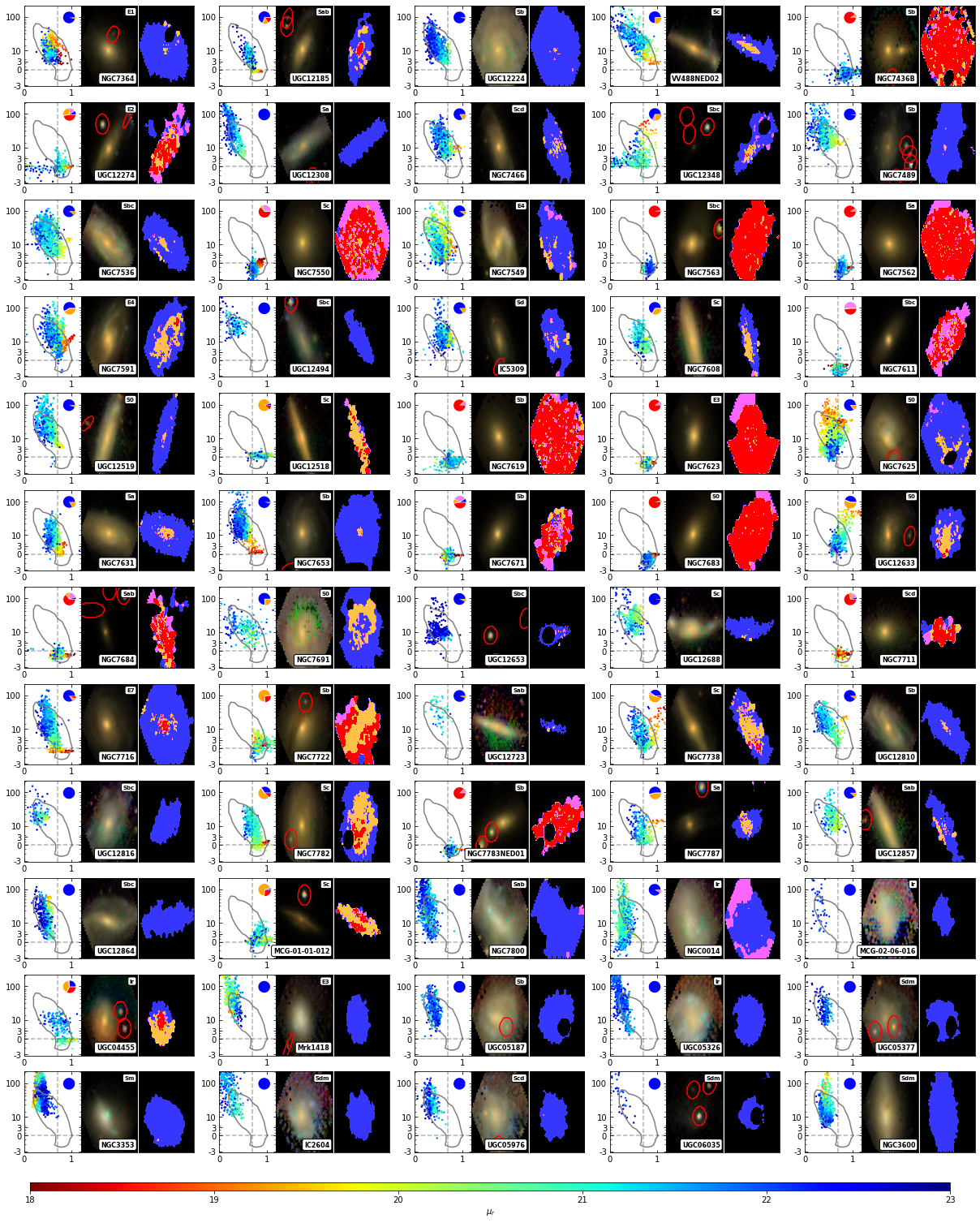}
    \contcaption{}
\end{figure*}

\begin{figure*}
    \centering
    \includegraphics[width=\linewidth]{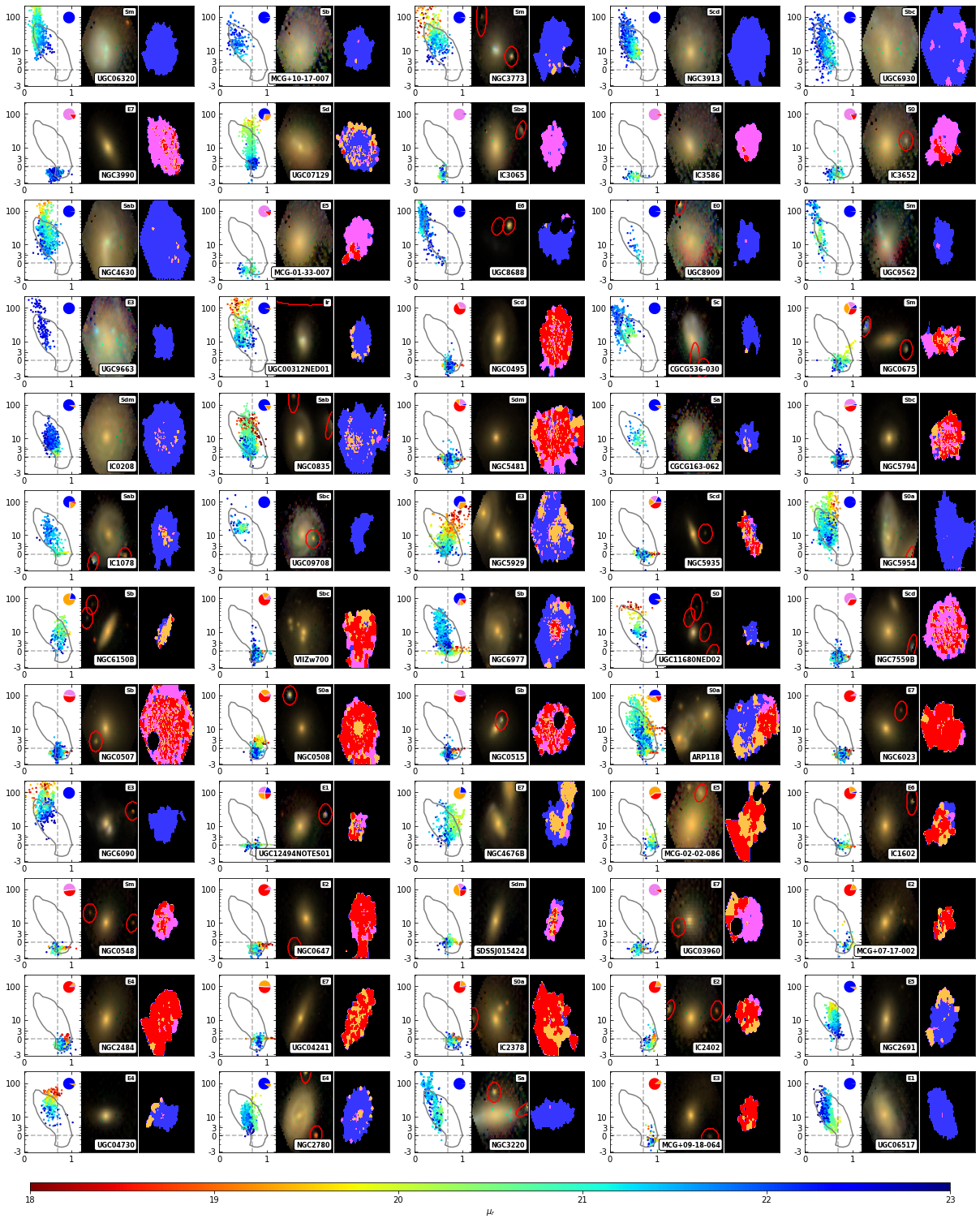}
    \contcaption{}
\end{figure*}

\begin{figure*}
    \centering
    \includegraphics[width=\linewidth]{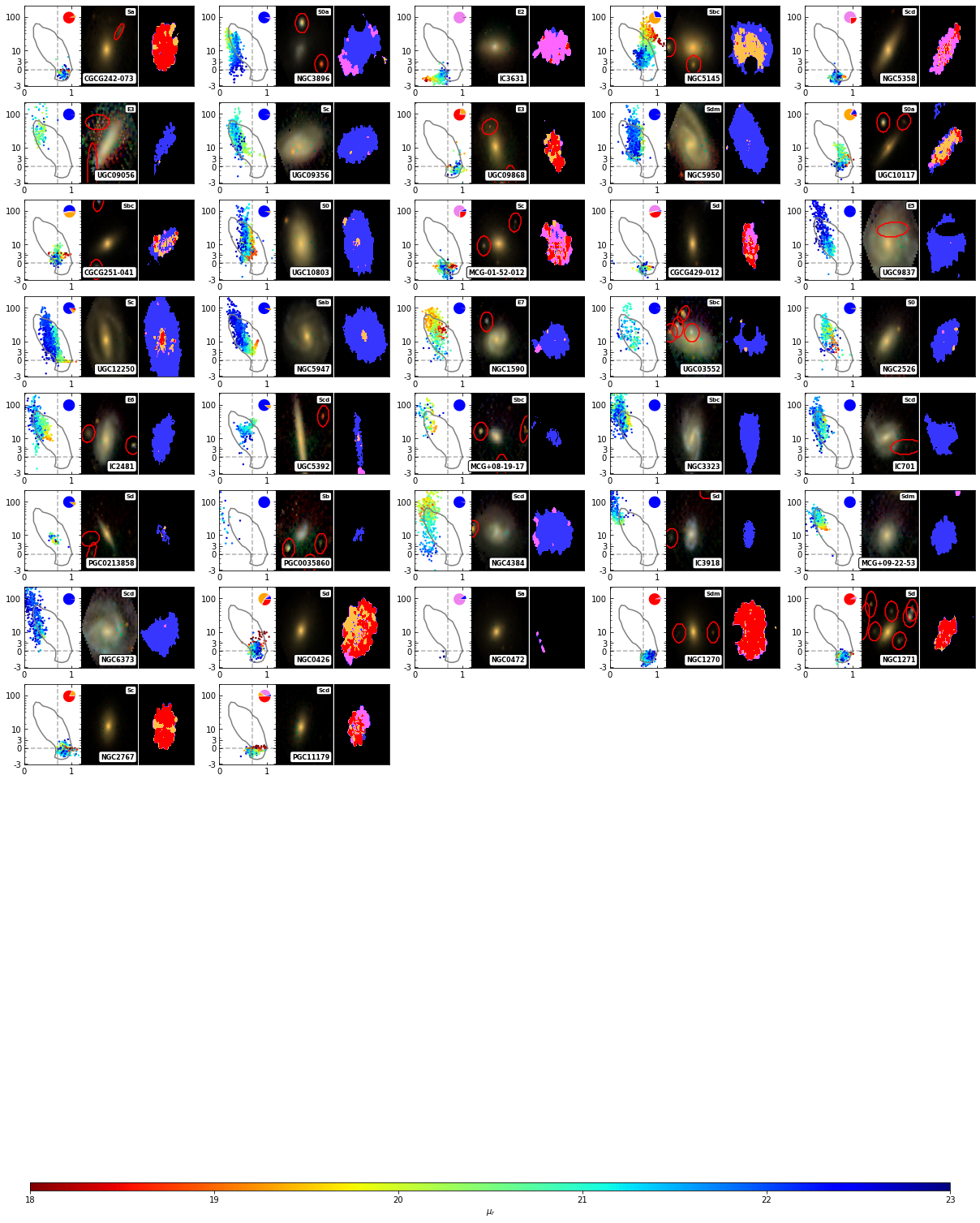}
    \contcaption{}
\end{figure*}
\bsp	
\label{lastpage}
\end{document}